\documentclass{lmcs} 


\usepackage{hyperref}
\theoremstyle{plain} 

\usepackage{graphicx}
\usepackage{amsmath, bm,tikz,amsthm}
\usepackage{amssymb}
\usepackage[inline]{enumitem}

\newcommand{\Swedge}[1]{\hat #1}
\newcommand{\Kpair}[1]{\widetilde #1}
\newtheorem{theorem}{Theorem}[section]
\newtheorem{remark}[theorem]{Remark}
\newtheorem{lemma}[theorem]{Lemma}
\newtheorem{example}[theorem]{Example}
\newtheorem{proposition}[theorem]{Proposition}
\newtheorem{definition}[theorem]{Definition}
\newtheorem{corollary}[theorem]{Corollary}
\newtheorem{question}[theorem]{Question}
\newtheorem{convention}[theorem]{Convention}
\newcommand{\mods}[1]{{#1}^{\rm M}}
\newcommand{\Mods}[1]{{#1}^{\rm m}}
\newcommand{\languni}{\lang_{\pd}^\forall}
\newcommand{\point}{\triangleright}
\newcommand{\putaway}[1]{}
\newcommand{\frm}[1]{{\mathcal{#1}}} 
\newcommand{\cls}[1]{{\mathbf{#1}}} 
\newcommand{\rel}[1]{R_{#1}} 
\newcommand{\power}[1]{2^{#1}}

\newcommand{\lft}{\mathfrak L}
\newcommand{\rgt}{\mathfrak R}

\newcommand{\lang}{{\mathsf L}}
\newcommand{\pd}{{\Diamond}}
\newcommand{\pointed}[1]{{\bm #1}}
\newcommand{\fsgm}{{\sc fgm}}
\newcommand{\fsgf}{{\sc fgf}}

\newcommand{\vaprhi}{\varphi}

\begin{document}

\title{Frame-validity games and  lower bounds on the complexity of modal axioms}
\titlecomment{{\lsuper*} This article extends \cite{AiML} presented at {\em Advances in Modal Logic} 2018.}

\author[P.~Balbiani]{Philippe Balbiani}	
\address{Institut de Recherche en Informatique de Toulouse\\
  CNRS --- Toulouse University}	
\email{\tt Philippe.Balbiani@irit.fr}  

\author[D.~Fern\'andez]{David Fern\'andez-Duque}	
\address{Department of Mathematics\\
  Ghent University}	
\email{\tt David.FernandezDuque@UGent.be}	

\author[A.~Herzig]{Andreas Herzig}	
\address{Institut de Recherche en Informatique de Toulouse\\
  CNRS --- Toulouse University}	
\email{\tt Andreas.Herzig@irit.fr}

\author[P.~Iliev]{Petar Iliev}	
\address{   Centre International de Math\'ematiques et d'Informatique de Toulouse\\
   Toulouse University}	
\email{\tt Petar.Iliev@irit.fr}





  \begin{abstract} 
We introduce frame-equivalence games tailored for reasoning about
 the size, modal depth, number of occurrences of symbols and number of
different propositional variables of modal formulae defining a given
frame-property. Using these games, we prove lower bounds on the above measures 
for a number of well-known modal axioms; what is more, for some of the axioms,
we show that they are optimal among the formulae defining the respective class
of frames.
\end{abstract}

\maketitle

\section{Introduction}

One of the key advantages of modal logics over first-order logic is that the former are often decidable. However, decidability is not sufficient for applications: efficiency plays a huge role in determining the usefulness of a formal system.
Typical measures of complexity revolve around problems such as satisfiability and model-checking, but the sometimes-overlooked {\em succinctness} plays a crucial role as well: there is little use in a {\sc ptime} logic if properties of interest can only be defined by exponentially large formulas.

The power of first-order logic and some of its
extensions to succinctly define graph properties has been investigated extensively \cite{ImmermanBook}, as that of the modal language and  natural extensions to define properties of relational {\em models} \cite{FernandezIliev,IlievCover}. 
In contrast, it seems that  the only  study of how succinctly  {\em frame} properties can be expressed in modal logic  is \cite{vakarelov}, where the question of how many different propositional variables are needed to modally define certain classes of Kripke frames is being considered.
To increase our understanding of the succinctness of  modal languages, we develop  in the present paper techniques for proving lower bounds on the complexity of modal formulas defining frame properties and apply them to some well-known classes of frames.

As usual, we say that a modal formula $\varphi$ defines a class $\cls{F}$ of frames if  
$\cls{F}$ exactly consists of the frames on which $\varphi$ is valid.
If a class of frames is definable by a 
modal formula, it is natural to ask how {\em complex} any such formula must be, where the complexity of a formula may be measured according to the total number of symbols, the modal depth, the number of occurrences of symbols of a certain type, or the number of different variables needed.

The techniques we will employ are based on {\em frame equivalence games,} closely related to model-equivalence games as appeared in~\cite{ijcai,succinctnessaijournal,hellaaiml}.
To demonstrate the applicability of the former to both first- and second-order semantic conditions, 
we prove that 
\begin{enumerate}
\item Every modal formula in a language with the universal box- and diamond-modalities that defines the class of graphs that are not 
$n$-colourable  must contain at least $\log_2 (n)$ different propositional variables and has size that is at least linear in $n$. 
As a counterpart to this lower bound, we provide a formula of a quasilinear length that contains $\log_2 (n)$ different propositional variables. 

As far as we are aware, previously known modal formulae defining this second-order property contain at least $n$ different propositional variables
and have size quadratic in $n$ (see, for example, \cite{regionspace}).
\item For each $m,n\geq 0$, the $(m,n)$-transfer axiom $\pd^m p\rightarrow \pd^n p$ is essentially the shortest modal formula defining
the first-order condition
\begin{equation}\label{eqGT}
\forall x \forall y (x R^m y \rightarrow  x R^n y),
\end{equation}
 where $R^j$ denotes the $j$-fold composition of $R$.
 Note that this result applies to the well-studied axioms defining transitivity, reflexivity, and density.
 \item The L\"{o}b  axiom $\Box(\Box p\rightarrow p) \rightarrow \Box p$ is essentially the
  shortest modal formula defining transitivity plus the second-order property of converse well-foundedness.
\item The formula $(p \vee \pd \pd p)\to \pd p$ is the shortest among those defining reflexivity plus transitivity.
\item The axiom $p\to \Box\pd p$ is the shortest modal formula that defines symmetry.
\end{enumerate}


The rest of the paper is organised as follows. 
The next section recalls some standard definitions. 
Section \ref{secMGames} describes formula-bound games on models and 
Section \ref{secFGames} turns them into games on frames. 
Section \ref{secNonColourable} provides shortest axioms for the $n$-colourability property. 
Section \ref{secGenTrans} provides a general result for transfer axioms.
Section \ref{secSFour} is about the $\mathsf{S4}$ axiom and 
Section \ref{secLob} is about the L\"{o}b axiom.
Section \ref{secConclusion} concludes. 
The appendix contains the proofs of sections \ref{secMGames} and 
an analysis of the symmetry axiom. 
The present paper extends a paper that was presented at AiML 2018.

\section{Technical preliminaries}\label{secTech}

Our formula size games are based on formulas in negation normal form, i.e., negations appear only in front of propositional symbols.
 Fix a countably infinite set of {\it propositional variables} $P=\{p_1, p_2, \ldots\}$, and let $\lang_\pd$ denote 
the uni-modal language that has as atomic formulas the {\em literals} $p$, $\overline p$ for each $p \in P$ as 
well as $\bot$, $\top$ and as primitive connectives $\vee$, $\wedge$, $\pd$, and $\Box $. 
The expressions $\neg \varphi$ and $\varphi \to \psi$ will be regarded as abbreviations defined using De Morgan's rules.
We will also be interested in the language $\languni$ that extends $\lang_\pd$ with the universal modalities $\exists$ and $\forall$.

As usual 
a frame is a pair $\frm A = (W_\frm A,R_\frm A)$ where $W_\frm A$ is a nonempty set and $R_\frm A \subseteq W_\frm A\times W_\frm A$; 
a {\em model based on $(W_\frm B,R_\frm B$)} is a tuple $\frm B = (W_\frm B,R_\frm B, V_\frm B)$ consisting of a frame equipped with a {\em valuation} $V_\frm B \colon W_\frm B \to 2^P$; and 
a {\em pointed model} is a tuple $\pointed c = (\frm C, c)$ consisting of a model $\frm C$ equipped with a designated point $c\in W_\frm C$. 
Pointed models will always be denoted by $\pointed a,\pointed b,\ldots$ and frames or models by $\frm A,\frm B,\ldots$
For a pointed model $ \pointed a = ({\frm A}, a)$, 
we denote by $\Box \pointed a$ the set $\{({\frm A}, b) : a \mathrel \rel{\frm A} b\}$, i.e., the set of all pointed models that are {\it successors}
of the pointed model $\pointed a$ along the relation $\rel{\frm A}$. Analogously, we use  $\forall \pointed a$ to denote the set $\{({\frm A}, b) :  b \in W_\frm A\}$.

Given $\varphi \in \languni$ and a pointed model $\pointed a$, we define $\pointed a \models \varphi$ according to standard Kripke semantics, and as usual if $\frm A$ is a model we write $\frm A \models \varphi$ if $(\frm A,a) \models \varphi $ for all $a\in W_\frm A$, and if $\frm A$ is a frame, $\frm A \models \varphi$ if $(\frm A,V) \models \vaprhi$ for every valuation $V$.
We use {\em structure} as an umbrella term to denote either a model, a frame, or a pointed model.
For a class of structures $\cls{A}$ and a formula $\varphi$, we write $\cls{A}\models \varphi$ when 
$ \frm X \models \varphi$ for all $ \frm X  \in \cls{A}$, and say that the formulae $\varphi$ and $\psi$ are {\it equivalent} on $\cls{A}$ when for all $\frm X \in \cls A$, $ \frm X   \models \varphi $ if and only if $ \frm X  \models \psi$.

Our goal is to develop techniques to establish when a formula $\varphi$ is of minimal complexity among those defining some class of frames.
Here {\em complexity} could mean many things: by a {\em complexity measure} (or just {\em measure}) we simply mean a function 
$\mu \colon \lang \to \mathbb N$, where $\lang$ is either $\lang_\pd$ or $\languni$.
We are interested in the following measures:
\begin{enumerate*}

\item the {\em length} of a formula $\vaprhi$, denoted $|\varphi|$ and defined as the number of nodes in its syntax tree (including leaves);

\item\label{itOcurr} the {\em number of ocurrences} of any connective,

\item the {\em modal depth,} and

\item the {\em number of variables.}

\end{enumerate*}

Note that these are a total of nine measures for $\lang_\pd$ and eleven for $\languni$, as each connective gives rise to its own measure in \eqref{itOcurr}.
We will show that several modal axioms of interest are minimal with respect to all of these measures simultaneously. To this end, given a set $\Gamma \subseteq \languni$ and $\varphi \in \Gamma$, we say that {\em $\varphi$ is absolutely minimal among $\Gamma$} if for all $\psi \in \Gamma$ and any of the  respective  measures $\mu$ described above, $\mu(\varphi) \leq \mu(\psi)$.

\section{A formula-bound game on models}\label{secMGames}

The game described below is the modal analogue of the formula-size game developed in the setting of first-order logic in \cite{adlerimmerman}. 
The general idea is that we have two competing players, {\it Hercules} and the {\it Hydra.}
Given two classes of pointed models $\cls{A}$ and $\cls{B}$ and $\lang \in \{\lang_\pd, \languni\}$, Hercules  is trying 
to show that there is a ``small'' $\lang$-formula $\varphi$ such that $\cls{A}\models \varphi$ but
$\cls{B}\models \neg \varphi$  whereas the Hydra is trying to show that any such 
$\varphi$ is ``big''.
The players move by adding and labelling nodes on a game-tree $\langle T,\preccurlyeq \rangle$. 
For our purposes a {\em tree} is a finite set partially ordered by some order $\preccurlyeq$ such that if $\eta \in T$ then ${\downarrow}\eta = \{ \nu : \nu \preccurlyeq \eta \}$ is linearly ordered; any set of the form ${\downarrow} \eta$ is a {\em branch} of $T$.
\begin{definition}\label{defSatisfactionGame}
The  $(\languni,\langle\cls{A},\cls{B}\rangle)$ {\em formula-complexity game on models} (denoted $(\languni,\langle\cls{A},\cls{B}\rangle)$-{\fsgm})
 is played by two players, Hercules and the Hydra, who construct a  game-tree $T$
 in such a way that each node $\eta\in T$ is  labelled
 with a pair  $\langle \lft (\eta), \rgt(\eta)\rangle$ of classes of pointed models and  either a literal or a symbol from  the set $\{\bot,\top,\vee,\wedge,\Diamond,\Box, \exists, \forall\}$ according to the rules below.

Any leaf $\eta$ 
can be declared either a {\em head} or a {\em stub}. Once $\eta$ has been declared 
a stub, no further moves can be played on it.  The construction of $T$ begins with a root labelled by
$\langle\cls{A},\cls{B}\rangle$ that is declared a head. 
Afterwards, the game continues as long as there is at least one head. In each turn, Hercules goes first by choosing a head $\eta$ labelled by $\langle \lft(\eta),\rgt(\eta)\rangle$. Hercules then plays the following moves, to which the Hydra possibly replies.
\smallskip

\noindent {\sc literal-move:} Hercules  chooses a literal $\iota$ such that  $\lft(\eta) \models\iota$  and  $\rgt(\eta)\models\neg  \iota$. 
The node $\eta$ is declared a stub and labelled with the symbol $\iota$.
\smallskip

\noindent {\sc $\bot$-move:} Hercules  can play this move only if $\lft(\eta)=\varnothing$. 
The node $\eta$ is declared a stub and labelled with the symbol $\bot$.
\smallskip

\noindent {\sc $\top$-move:} Hercules  can play this move only if $\rgt(\eta)=\varnothing$. 
The node $\eta$ is declared a stub and labelled with the symbol $\top$.
\smallskip

\noindent {\sc $\vee$-move:} Hercules labels $\eta$ with the symbol $\vee$ and chooses two sets  
$\cls{L}_1,\cls{L}_2 \subseteq\cls{L}$ such that
  $\lft(\eta) =\cls{L}_1 \cup \cls{L}_2$. Two new heads, labelled by $\langle\cls{L}_1,\rgt(\eta)\rangle$ and 
  $ \langle\cls{L}_2,\rgt(\eta)\rangle$, are added to $T$ as daughters of $\eta$.
\smallskip

\noindent {\sc $\wedge$-move:} Dual to the $\vee$-move, except that in this case Hercules chooses $\cls R_1$, $\cls R_2$ such that $\cls R_1 \cup \cls R_2 = \rgt(\eta)$.
\smallskip

\noindent {\sc $\Diamond$-move:} Hercules labels $\eta$ with the symbol $\Diamond$ and, for each pointed 
model $\pointed l \in \lft(\eta)$, chooses a pointed model from $\Box \pointed l$; if for some  $\pointed l \in \lft(\eta)$ we have $\Box \pointed l=\varnothing$,
 Hercules cannot play this move. All these new pointed models  are collected in 
the set $\cls{L}_1$.
For each pointed model $\pointed r \in \rgt(\eta)$, the Hydra  replies
 by picking a subset of $\Box \pointed r$.
All the pointed models chosen by the Hydra are 
collected in the class $\cls{R}_1$.\footnote{
In particular, if $\Box \pointed r=\varnothing$ 
for some $\pointed r \in \rgt(\eta)$ 
then $\cls{R}_1 = \emptyset$, i.e., the Hydra does not 
add anything to $\cls{R}_1$. 
For example, when 
$\lft(\eta)$ is the set of all serial models and
$\rgt(\eta)$ contains a model built on the irreflexive singleton frame $\langle \{w\}, \emptyset \rangle$ 
then a $\Diamond$-move on $\eta$ results in a new head labelled $\langle \lft(\eta), \emptyset \rangle$. 
} 
A new head labelled by $\langle\cls{L}_1, \cls{R}_1\rangle$  is added as a daughter to $\eta$.
\smallskip

\noindent {\sc $\Box$-move:} Dual to the $\pd$-move, except that Hercules first chooses a successor for each $\pointed r \in \cls{R}$ and Hydra chooses her successors for frames in $\cls L$.
\smallskip

\noindent {\sc $\exists$-move:} Hercules labels $\eta$ with the symbol $\exists$ and, for each pointed 
model $\pointed l \in \rgt(\eta)$,  he chooses a pointed model from $\forall \pointed l$. All these new pointed models  are collected in 
the set $\cls{L}_1$.
For each pointed model $\pointed r \in \rgt(\eta)$, the Hydra  replies
 by picking a subset of $\forall \pointed r$.
All the pointed models chosen by the Hydra are 
collected in the class $\cls{R}_1$. 
A new head labelled by $\langle\cls{L}_1, \cls{R}_1\rangle$  is added as a daughter to $\eta$.
\smallskip

\noindent {\sc $\forall$-move:} Dual to the $\exists$-move, except that Hercules first chooses a successor for each $\pointed r \in \cls{R}$ and Hydra chooses her successors for frames in $\cls L$.
\smallskip

\noindent The $(\languni, \langle\cls{A},\cls{B}\rangle)$-{\fsgm} concludes when there are no heads and we say  in this case that  $T$ is a {\em closed game tree}.
\end{definition}

Note that the Hydra has no restrictions on the number of pointed models she chooses on modal moves; in fact, she can choose all of them, and it is often convenient to assume that she always does so.
To be precise, say that the Hydra {\em plays greedily} if
\begin{enumerate*}
\item 
whenever Hercules makes a $\pd$-move on a node $\eta$ and a new node $\eta'$ is added then $\rgt ({\eta'}) = \bigcup_{\pointed r \in \rgt (\eta)} \Box \pointed r$, and similarly 
\item
whenever Hercules makes a $\Box$-move on a node $\eta$ and a new node $\eta'$ is added then $\lft ({\eta'}) = \bigcup_{\pointed l \in \lft (\eta) } \Box \pointed l$,
\item 
analogously for $\exists$- and $\forall$-moves.
\end{enumerate*}

The $(\languni, \langle\cls{A},\cls{B}\rangle)$-{\fsgm} can be used to give lower bounds on the length of $\languni$-formulae defining a given property; if we are interested
in the length of formulae in the sub-language $\lang_{\pd}$ of $\languni$ that does not have $\exists$ and $\forall$ operators, we simply do not allow the corresponding $\exists$ and $\forall$-moves and this new game is denoted $(\lang_{\pd}, \langle\cls{A},\cls{B}\rangle)$-{\fsgm}~\cite{ijcai,succinctnessaijournal,hellaaiml}.
 Here we will generalize these games to show that they can be used to give lower bounds on any complexity measure. For this, we need to view game-trees as formulae.

\begin{definition}
Given a closed $(\languni, \langle\cls{A},\cls{B}\rangle)$-{\fsgm} tree $T$, we define $\psi_T \in \languni$ to be the unique formula whose syntax tree is given by $T$.
\end{definition}

\noindent Formally speaking, $\psi_T$ is defined by recursion on $T$ starting from  leaves: if $T$ is a single leaf then it must be labelled by a literal $\iota$, or by $\bot$, or by $\top$, so we respectively set $\psi_T = \iota$, or $\psi_T = \bot$, or $\psi_T = \top$; if $T$ has a root $\eta$ labelled by $\vee$, then $\eta$ has two daughters $\eta_1$, $\eta_2$. Letting $T_1$, $T_2$ be the respective generated subtrees, we define $\psi_T = \psi_{T_1}\vee \psi_{T_2}$.
The cases for $\wedge$, $\Diamond$,  $\Box$, $\exists$, and $\forall$ are all analogous.
Then, given a complexity measure $\mu$, we extend the domain of $\mu$ to include the set of closed game trees by defining $\mu(T) = \mu (\psi_T)$.

If $\lang \in \{ \lang_\pd, \languni \}$, $m \in \mathbb N$, $\cls A$, $\cls B$ are classes of models, and $\mu \colon \lang \to \mathbb N$ a complexity measure (including but not restricted to the four measures that we have defined in Section \ref{secTech}), 
we say that Hercules has a {\em winning strategy for the $(\lang, \langle\cls{A},\cls{B}\rangle)$-{\fsgm} 
with $\mu $ below $ m$} if Hercules has a strategy so that no matter how Hydra plays, the game terminates in finite time with a closed tree $T$ so that $\mu(T) < m$.

\begin{theorem}\label{thrm: satisfactionGames}
Let $\lang \in \{ \lang_\pd, \languni \}$, $\cls A$, $\cls B$ be classes of models, $\mu \colon \lang \to \mathbb N$ any complexity measure, and $m \in \mathbb N$. Then the following are equivalent:
\begin{enumerate}

\item Hercules has a winning strategy for the $(\lang , \langle\cls{A},\cls{B}\rangle)$-{\fsgm} with $\mu$ below $m$;

\item there is an $\lang $-formula $\varphi$ with $\mu(\varphi) < m$ 
 and $\cls{A}\models\varphi $ whereas $\cls{B}\models\neg\varphi$.
 \end{enumerate}
\end{theorem}

\noindent 
We defer the proof of Theorem \ref{thrm: satisfactionGames} to Appendix \ref{apSGames}, where we also establish some useful properties of the formula-complexity game. However, we remark that the proof is essentially the same as that of the special case where $\mu(\varphi ) = |\varphi|$, which can be found in any of \cite{ijcai,succinctnessaijournal,hellaaiml}.
We will also use the following easy consequence of Theorem \ref{thrm: satisfactionGames}.
We assume familiarity with bisimulations \cite{Chagrov1997}.

\begin{corollary}\label{lemHercLose}
Let $\lang \in \{ \lang_\pd, \languni \}$, $\cls A$ and $\cls B$ be classes of pointed models such that there are
$\pointed a \in \cls A$ and $\pointed b \in \cls B$ with $\pointed a$ $\lang$-bisimilar to $\pointed b$.
For all complexity measures $\mu$ and for all non-negative integers $m$, Hercules has no winning strategy for the $(\lang , \langle\cls{A},\cls{B}\rangle)$-{\fsgm} with $\mu$ below $m$.
\end{corollary}

\section{A formula-complexity game on frames}\label{secFGames}

We  develop an analogous game to the one above that
is played on frames instead of models in order to reason about the ``resources'' 
needed to modally define properties of frames with $\lang_{\pd}$- or $\languni$-formulas.

\begin{definition}
Let $\cls A$, $\cls B$ be classes of frames. The $( \languni,\langle \cls{A},\cls{B}\rangle)$ for\-mula-complexity game on frames (denoted $( \languni,\langle \cls{A},\cls{B}\rangle)$-{\fsgf}) is played  by Hercules and the Hydra as follows.
\medskip

\noindent{\sc Hercules Selects Models:}
For each $\frm{B}\in \cls{B}$ Hercules chooses a model $\mods{\frm B}$ based on $\frm B$ and a point $\triangleright_\frm B \in W_\mathcal B$ and then sets $\Mods{\cls B} = \{(\mods{\frm B},\point_\frm B):\frm B \in \cls B\}$.
\medskip

\noindent{\sc The Hydra Selects Models:} The Hydra replies by choosing a class of pointed models $\Mods{\cls A}$ of the form $(\frm A,V,a)$ with $\frm A \in \cls A$.
\medskip
 
\noindent{\sc Formula Game on Models:} Hercules and the Hydra play the
 $( \languni, \langle \Mods{\cls A}, \Mods{\cls B} \rangle)$-{\fsgm}.
 \medskip

\noindent The game tree assigned to a match of the $( \languni,\langle \cls{A},\cls{B}\rangle)$-{\fsgf} is the game tree of the subsequent $( \languni, \langle \Mods {\cls{A}}, \Mods{\cls{B}} \rangle)$-{\fsgm}.  
As before, if we are interested in the length of $\lang_{\pd}$-formulae we do not allow
$\exists$- or $\forall$-moves, and the resulting game is denoted the $( \lang_{\pd}, \langle \Mods {\cls{A}}, \Mods{\cls{B}} \rangle)$-{\fsgf}.

\end{definition} 

\begin{remark}
The Hydra is free to assign as many models as she wants to each $\frm A \in \cls A$, even no model at all.
We say that the Hydra plays {\em functionally} if she chooses $\Mods{\cls A}$ so that for each $\frm A \in \cls A$ there is exactly one pointed model $(\mods{\frm A},\point_\frm A) \in \Mods{\cls A} $ with $\mods{\frm A}$ based on $\frm A$.
In this text the Hydra will often play functionally.
\end{remark}

As was the case for the {\fsgm}, for $\lang \in \{\lang_\pd, \languni \}$, $m \in \mathbb N$, classes of frames $\cls A$, $\cls B$, and $\mu \colon \lang \to \mathbb N$ a complexity measure, Hercules has a {\em winning strategy for the $(\languni, \langle \cls{A},\cls{B}\rangle)$-{\fsgf} with $\mu $ below $ m$} if 
Hercules has a strategy such that, no matter how Hydra plays, the game terminates in finite time with a closed tree $T$ so that $\mu(T) < m$.

\begin{theorem}\label{thrm: validityGames}
Let $\lang \in \{\lang_\pd, \languni \}$, $\cls A$, $\cls B$ be classes of frames, $\mu $ any complexity measure, and $m \in \mathbb N$. Then, the following are equivalent:
\begin{enumerate}

\item\label{itValGOne} Hercules has a winning strategy for the $(\lang, \langle\cls{A},\cls{B}\rangle)$-{\fsgf} with $\mu$ below $m$;

\item\label{itValGTwo} there is an $\lang$-formula $\varphi$ with $\mu(\varphi) < m$ 
 that is valid on every frame of $\cls{A}$ and non-valid on every frame of $\cls{B}$.
 \end{enumerate}
\end{theorem}
\proof {\sc \eqref{itThmSatTwo}  implies \eqref{itValGOne}.} Let $\varphi$ be an $\lang$-formula with $\mu(\varphi) < m$ that is valid on all frames in $\cls{A}$ and not valid on any frame in $\cls{B}$.
 For each $\frm B \in \cls B$, Hercules can choose a pointed model $\mods{\frm B} = (\frm B,V,b)
$ based on $\frm B$ so that $\mods{\frm B} \not \models \varphi$.
The Hydra then responds with some set of pointed models $\Mods{\cls A}$; since $\varphi$ is valid on $\cls A$, for all $\frm A\in \Mods{\cls A}$ we have  $\frm A \models \varphi$.
By Theorem~\ref{thrm: satisfactionGames}, it follows that Hercules has a winning strategy with $\mu$ below $m$ for the  $(\lang, \langle \Mods{\cls{A}}, \mods{\cls{B}} \rangle)$-{\fsgm} and thus for $(\lang,\langle \cls{A},\cls{B}\rangle)$-{\fsgf}.
\smallskip

\noindent {\sc  \eqref{itThmSatOne}  implies \eqref{itValGTwo}.} Now assume that Hercules has such a strategy, and that he chooses $\Mods{\cls B}$ according to this strategy.
Then Hydra {\em opens greedily} by choosing every pointed model based on a frame in $\cls A$; in other words, she sets $\Mods{\cls A}$ to be the set of all $(\frm A,V,a)$ with $\frm A\in \cls A$, $V$ a valuation on $\frm A$ and $a\in W_\frm A$.

Assume that the Hydra opens greedily.
By playing according to his strategy, Hercules can win the $(\Mods{\cls A},\Mods{\cls B})$-{\fsgm} with a closed game tree $T$ such that $\mu(T) < m$; but this is only possible if his sub-strategy for the $(\Mods{\cls A},\Mods{\cls B})$-{\fsgm} is a winning strategy with $\mu$ below $m$.
Thus by Theorem~\ref{thrm: satisfactionGames}, there is a $\lang$-formula $\varphi$ with $\mu(\varphi) < m$ such that $\Mods{\cls{A}} \models \varphi$
and $\Mods{\cls{B}} \models \neg\varphi$.
Since Hercules chose one pointed model for each $\frm B\in \cls B$ it follows that  $\varphi$ is not valid in any frame in $\cls B$, while since Hydra chose all possible pointed models it follows that $\frm A \models \varphi$.
\endproof

In the next sections, we apply our formula-complexity games to prove lower bounds on the complexity  of some modal axioms. 
  For ease of understanding, we define the pointed models employed in our proofs using figures.
We follow the convention that such pointed models consist of the relevant Kripke model and a point that is
denoted by the `$\triangleright$' sign next to it.

\section{The Non-Colourability Property}\label{secNonColourable}

For a natural number $n\geq 1$, let us consider the property of a graph not being $n$-colourable, i.e, the set of its vertices cannot be partitioned
in at most $n$ equivalence classes so that no two vertices sharing the same edge are in the same equivalence class. 
This property is modally definable with the help of the universal modalities $\exists$ and $\forall$. A natural way of finding a defining modal formula 
is to reason as follows. To encode the $n$ colours, we can use the propositional symbols $p_1, \ldots, p_n$, respectively.
Then,  we  write a $\languni$-formula with the help of $\exists$ and $\forall$  that says ``if every node of the graph is coloured with
exactly one colour, then there are two edge-related nodes that have the same colour''. Formally,
$$\forall \ \Big ( (p_1\vee\ldots\vee p_n) \wedge
\big ( \bigwedge_{1\leq i< j\leq n} \lnot (p_i\wedge p_j) \big) \Big )
\rightarrow
\exists \ \big (\bigvee_{1\leq i\leq n}(p_i\wedge\pd p_i) \big ). $$
A version of the above formula can be found in~\cite{regionspace}.
Because  of the subformula 
$\bigwedge\limits_{1\leq i< j\leq n} \lnot (p_i\wedge p_j)$, 
the length of the whole
formula is quadratic in $n$. We show below that we can do much better and find a formula of a quasilinear length and with exponentially smaller number of variables that expresses the non-colourability property.

Recall that $P$ denotes the set of propositional variables.
For any natural number $k \geq 1$, let $P_k\subset P$ be the subset of $P$ containing only the first  $k$ variables in $P$.

\begin{definition}\label{def:noncol}
We define a sequence of formulas $(\varphi_n)_{n=1}^\infty$ as follows.

For $n=1$ we set $\varphi_1 = \exists \pd \top$.
If $n\geq 2$, let $k = \lceil \log_2 n \rceil $ (so that $2^{k-1}< n\leq 2^k$).
Fix an enumeration $\{S_1, \ldots, S_{2^k}\}$ of $\power {P_k}$, and to every $E \subseteq P_k$, associate
an elementary conjunction $\Swedge E$ defined by
\[\Swedge E = \bigwedge_{p\in E}p \wedge \bigwedge_{p\in P_k\setminus E} \overline{p}.\]
Then, let $\varphi_n$ be the formula 
\[
\exists \ \Big (\bigvee_{1\leq i\leq n} (\Swedge S_i \wedge \pd \Swedge S_i  )\vee \bigvee_{{n+1}\leq j\leq 2^k}
\Swedge S_j  \Big ).
\]
\end{definition} 

For example, for $n=2$ we have $k=1$, so $\varphi_2$ is 
$ \exists( (\overline p \wedge \pd \overline p) \vee (p \wedge \pd p) )$.
Since $2^k < 2n$ and each $S_i$ contains less than $ \log_2 ( n ) + 1 $ propositional variables, it is easily seen that there are less than $2\cdot 2n \big (\log_2 ( n ) + 1\big) $ 
occurrences of propositional variables in $\varphi_n$, and similarly that the lengths of $\varphi_n$ are bounded 
from above by a function in $O \big (n \log_2 (n) \big )$.
Moreover, the formulas $\varphi_n$ characterize non-$n$-colourability.
Below, note that directed graphs are just Kripke frames, hence we can speak of validity of a formula on a directed graph.
We will moreover regard non-directed graphs as directed graphs with a symmetric edge relation.

\begin{proposition} For any graph $\mathcal G$, $\varphi_n$ is valid on $\mathcal G$ iff $\mathcal G$ is not $n$-colourable.
\end{proposition}

\proof
 We begin by showing that if ${\mathcal G}$ is $n$-colourable then $\varphi_n$ is not valid in $\mathcal G$. Suppose that $W_\mathcal G$ can be partitioned in $n$ equivalence classes $C_1,\ldots,C_n$ so that no two vertices sharing the same edge belong to the
same equivalence class.
Recall that $\{S_1, \ldots, S_{2^k}\}$ is an enumeration of all subsets of $P_k$.
We define a valuation on $\mathcal G$ by setting $p\in V(w)$ if and only if for the unique $i$ such that $w\in C_i$ we have that $p \in S_i.$
It is immediate that the negation of $\varphi_n$, 
$$\forall \ \big (\bigwedge_{1\leq i\leq n} (\Swedge S_i \rightarrow\neg \pd \Swedge S_i  )\wedge \bigwedge_{{n+1}\leq j\leq 2^k} \neg \Swedge S_i \big ), $$
is true in the model $(\mathcal G,V)$.

Conversely, assume that $\varphi_n$ is not valid in $\mathcal G$. Therefore, there is a valuation $V$ such 
that $\neg\varphi_n$ holds, i.e., 
$$\forall(\bigwedge _ {1\leq i\leq n} (\Swedge S_i \rightarrow\neg \pd \Swedge S_i )\wedge \bigwedge_{{n+1}\leq j\leq 2^k}\neg \Swedge S_j) $$
is true in the resulting Kripke model $(\mathcal G,V)$. 
It is easily seen that this implies that $\mathcal G$ can be $n$-coloured by defining, for $i\in[1,n]$, $C_i$ to be the set of all $ w\in W_\mathcal G$ such that for all $p\in P_k$, $ p \in V(w)$ if and only if $p \in S_i$.
\endproof

In the rest of this section, we establish a linear lower bound on the size of any $\languni$-formula that defines
non-$n$-colourability.
Since the formulas $\varphi_n$ are quasi-linear on $n$, we leave the question of a sharp lower bound open.
 
\begin{theorem}\label{thm:noncol}  For any natural number $n\geq 2$, 
any $\languni$-formula $\varphi$ that defines the property of a graph being non-$n$-colourable contains at least 
$\lceil \log_2 (n)\rceil$ different propositional symbols, at least one occurrence of $\exists$, and has size at least $n$. 
\end{theorem}

We begin the proof of Theorem~\ref{thm:noncol} by proving the bound on the number of variables.
Recall that the complete graph on $n$ nodes, usually denoted ${\frm K}_n$, is an undirected, irreflexive graph with $n$ vertices in which every pair of distinct vertices is connected by an edge.
Since Kripke semantics are based on directed graphs, we will regard ${\frm K}_n = (W_n,R_n)$ as a directed graph, albeit with a symmetric relation, so that $w\mathrel R_n v$ if and only if $w \not = v$.
Clearly, every ${\frm K}_n$ is $n$-colourable. 

For $n\geq 1$, we let ${\Kpair{{\frm K}}}_n$ be a graph that consists of two disjoint copies of
${\frm K}_n$ so that only one of the copies of  ${\frm K}_n$ contains exactly one  reflexive node.
The graph ${\Kpair{{\frm K}}}_n$ is formally defined as follows.

\begin{definition}
Let $n\geq 1$ and fix $s\in W_n$. We define ${\Kpair{{\frm K}}}_n = (\Kpair W_n, \Kpair R_n)$, where $\Kpair W_n = W_n \times \{{\rm i},{\rm r}\}$ and $(w,x) \mathrel{\Kpair R_n} (v,y)$ if and only if either $w\not=x$ and $x=y$, or $w=v=s$ and $x=y={\rm r}$.

We call $W_n \times\{{\rm i}\}$ the the {\em irreflexive component}
of ${\Kpair{{\frm K}}}_n$ and $W_n \times\{{\rm r}\}$
the {\em reflexive component} of ${\Kpair{{\frm K}}}_n$. 
\end{definition}
 
\begin{example}
The graph ${\Kpair{{\frm K}}}_3$ is shown in Figure~\ref{fig:k3hat}.
\begin{figure} [htbp]
\begin{center}
\begin{tikzpicture}
[
help/.style={circle,draw=white!100,fill=white!100,thick, inner sep=0pt,minimum size=1mm},
white/.style={circle,draw=black!100,fill=white!100,thick, inner sep=0pt,minimum size=2mm},
black/.style={circle,draw=black!100,fill=black!100,thick, inner sep=0pt,minimum size=2mm},
rect/.style={rectangle,draw=black!100,fill=black!20,thick, inner sep=0pt,minimum size=2mm},
wrect/.style={rectangle,draw=black!100,fill=white!100,thick, inner sep=0pt,minimum size=2mm},
brect/.style={rectangle,draw=black!100,fill=black!100,thick, inner sep=0pt,minimum size=2mm},
grey/.style={circle,draw=black!100,fill=black!20,thick, inner sep=0pt,minimum size=2mm},
]


\node at (1,1) [white](nonreflLeft) [ ]{};
\node at (1.5,1.5) [white](nonreflMid) [ ]{};
\node at (2,1) [white](nonreflRight) [ ]{};

\node at (2.5,1) [white](reflLeft) [ ]{};
\node at (3,1.5) [white](reflMid) [ ]{};
\node at (3.5,1) [white](reflRight) [ ]{};

\begin{scope}[>=stealth, auto]
\draw [-] (nonreflLeft) to  (nonreflMid);
\draw [-] (nonreflLeft) to  (nonreflRight);
\draw [-] (nonreflMid) to  (nonreflRight);

\draw [-] (reflLeft) to  (reflMid);
\draw [-] (reflLeft) to  (reflRight);
\draw [-] (reflMid) to  (reflRight);

\draw [->] (reflMid) [out=50, in=130, loop]to  (reflMid);

\end{scope}

\end{tikzpicture}
\end{center}
\caption{The graph ${\Kpair{{\frm K}}}_3$.}
\label{fig:k3hat}

\end{figure}

\end{example}
 Obviously, due to the presence of the reflexive point, any ${\Kpair{{\frm K}}}_n$
 is a non-$n$-colourable graph.

\begin{lemma}\label{lemma:nonColBisimilar} For every valuation $V$ on $ {\frm K}_n  $ there is a valuation $\Kpair V$ on $\Kpair{{\frm K}}_n$ such that $(\frm K_n,V)$ is $\languni$-bisimilar to $(\Kpair{{\frm K}}_n, \Kpair V)$.
\end{lemma}

\proof Let us fix a pair of vertices $u$, $s$ in ${\frm K}_n$ that satisfy the same propositional variables. 
The model $\mods{{\Kpair{{\frm K}}}_n}$ then 
consists of two disjoint copies of the model $\mods{\frm K}_n$ but in one of the copies one of the points $u$ or $s$
is reflexive. 
It is easy to see  that  $\mods{\frm K}_n$ is  $\languni$-bisimilar to  $\mods{{\Kpair{{\frm K}}}_n}$.
\endproof

\begin{example}
The bisimilar  models $( {{\Kpair{{\frm K}}}_3},\Kpair V )$   and   $({\frm K}_3,V)$ are  shown in Figure~\ref{fig:k3hatModels} on the
left and right of the dotted line, respectively. All black nodes satisfy the same propositional variables.
\begin{figure} [htbp]
\begin{center}
\begin{tikzpicture}
[
help/.style={circle,draw=white!100,fill=white!100,thick, inner sep=0pt,minimum size=1mm},
white/.style={circle,draw=black!100,fill=white!100,thick, inner sep=0pt,minimum size=2mm},
black/.style={circle,draw=black!100,fill=black!100,thick, inner sep=0pt,minimum size=2mm},
rect/.style={rectangle,draw=black!100,fill=black!20,thick, inner sep=0pt,minimum size=2mm},
wrect/.style={rectangle,draw=black!100,fill=white!100,thick, inner sep=0pt,minimum size=2mm},
brect/.style={rectangle,draw=black!100,fill=black!100,thick, inner sep=0pt,minimum size=2mm},
grey/.style={circle,draw=black!100,fill=black!20,thick, inner sep=0pt,minimum size=2mm},
]


\node at (1,1) [black](nonreflLeft) [ ]{};
\node at (1.5,1.5) [black](nonreflMid) [ ]{};
\node at (2,1) [white](nonreflRight) [ ]{};

\node at (2.5,1) [black](reflLeft) [ ]{};
\node at (3,1.5) [black](reflMid) [ ]{};
\node at (3.5,1) [white](reflRight) [ ]{};

\node at ( 4,2) [help](helpUpDotted) []{};
\node at ( 4,1) [help](helpDownDotted) []{};

\begin{scope}[>=stealth, auto]
\draw [-] (nonreflLeft) to  (nonreflMid);
\draw [-] (nonreflLeft) to  (nonreflRight);
\draw [-] (nonreflMid) to  (nonreflRight);

\draw [-] (reflLeft) to  (reflMid);
\draw [-] (reflLeft) to  (reflRight);
\draw [-] (reflMid) to  (reflRight);

\draw [->] (reflMid) [out=50, in=130, loop]to  (reflMid);

\draw [ dotted] (helpUpDotted) to  (helpDownDotted);
\end{scope}

\node at (4.5,1) [black](kLeft) [ ]{};
\node at (5,1.5) [black](kMid) [ ]{};
\node at (5.5,1) [white](kRight) [ ]{};

\begin{scope}[>=stealth, auto]
\draw [-] (kLeft) to  (kMid);
\draw [-] (kLeft) to  (kRight);
\draw [-] (kMid) to  (kRight);

\end{scope}
\end{tikzpicture}

\end{center}
\caption{The model $(\mods{{\Kpair{{\frm K}}}_3},\Kpair V)$ (left), where $(\mods{\frm K}_3,V)$ (right) is such that two points share the same valuation.}
\label{fig:k3hatModels}
\end{figure}
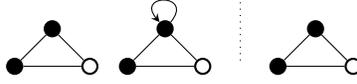

\end{example}


\begin{proposition}\label{prop:noncolvar}  For any natural number 
$n\geq 2$, any  $\languni$-formula $\varphi$ that defines the property of a graph being non-$n$-colourable contains at least 
$\lceil \log_2(n)\rceil$ different propositional symbols.
\end{proposition}

\proof Let $k= \lceil \log_2(n)\rceil$, $\ell <k$, and $\psi \in \languni$ a formula containing only  $\ell$ different propositional 
variables, say $p_1,\ldots, p_\ell$.
We are going to show that this formula is either valid on $\frm K_n$ or not valid on ${{\Kpair{{\frm K}}}_n}$,
and hence $\psi$ does not define the property of not being $n$-colourable.

Assume that $\varphi$ is not valid on $\frm K_n$, and let $V$ be a valuation so that $(\frm K_n, V) \not \models \varphi$.
The assumption that $\ell < k$ implies that $n>2^\ell$ and, therefore,
there are at least two different nodes $u$ and $v$ in ${\frm K}_n$ that satisfy the same  subset of 
$\{p_1,\ldots, p_\ell\}$. 
Applying Lemma~\ref{lemma:nonColBisimilar}, we obtain a valuation $\Kpair V$ on $\Kpair {{\frm K}}_n$ such that $(\Kpair {{\frm K}}_n, \Kpair V)$ is $\languni$-bisimilar to $(\frm K_n,V)$, so there must be a point in  $\mods{{\Kpair{{\frm K}}}_n}$ that falsifies $\psi$.
\endproof

This establishes the lower bound on the number of variables of Theorem \ref{thm:noncol}.
For the rest of the properties we will consider a formula game on frames.
Let us fix an $n\geq 1$  and consider a  $( \languni,\langle \cls{A},\cls{B}\rangle)$-{\fsgf} where
$\cls{A}=\{{\Kpair{{\frm K}}}_n\}$ and $\cls{B}=\{\frm{K}_n\}$.
Clearly, the formula $\varphi_n$ of Definition \ref{def:noncol} is valid on the frame in $\cls{A}$ and not valid on the frame in $\cls{B}$.
Below we will detail the strategy that Hercules must follow if the Hydra plays greedily and he wishes to win the game. We begin with his selection of models.\\

\noindent{\sc selection of the models on the right:} It follows from 
Lemma~\ref{lemma:nonColBisimilar} that if Hercules wants to win the subsequent {\fsgm},
he must choose his model  $\frm B = ({\frm K}_n,V)$ so that any two different vertices of $\frm B$
satisfy different sets of literals. 
Let the singleton set $\Mods{\cls{B}}$ contain the pointed model $(\frm B, w)$ chosen by 
Hercules, where $w\in W_n$ is arbitrary.\\ 

\noindent{\sc selection of the models on the left:} The Hydra constructs a set $\Mods{\cls{A}}_n$ of $n$ different  pointed models  based on ${\Kpair{{\frm K}}}_n$ as follows.
Intuitively, for each $w\in W_n$ she will construct a model $\frm A_w$ consisting of two copies of $\frm B$, where in the second copy the reflexive point satisfies the same propositional variables as $w$.

More formally, let $(s,{\rm r})$ be the unique reflexive point of ${\Kpair{{\frm K}}}_n$.
For each $w\in W_n$, let $\pi_w$ be a permutation of $W_n$ such that $\pi_w(s) = w$.
Then, we define $V_w(u,x) = V(\pi_w(u))$, and define $\frm A_w = ({\Kpair{{\frm K}}}_n,V_w)$.
Finally, we set $\cls A_n = \{\frm A_w : w\in W_n\}$.

\begin{convention}
We will henceforth notationally identify a vertex $w\in W_n$ with the set of propositional variables $V(w)$; note that, since Hercules assigns different valuations to different points, a set of variables $E$ can name at most one vertex.
Similarly we will denote a vertex $(v,x)$ of $\frm A_w$ by $E^x$ if $E = V_w(v,x)$. 
\end{convention}

For example, we may write $(\frm A_E,S^x)$ instead of $(\frm A_w , ( v,x ) )$ if $E = V(w)$ and $S=V_w ( w ,x)$, or write $E \in W_n$ to indicate that $E=V(w)$ for some $w\in W_n$.
Note that there is a slight ambiguity in the notation since, strictly speaking, $S^x$ might denote a different point in $\frm A_E$ than it does in $\frm A_{E'}$; however, this slight ambiguity is innocuous (and can in fact be eliminated altogether by suitably permuting the elements of each domain).

\begin{example}
The classes of pointed models $\Mods{\cls{A}}_3$ and $\Mods{\cls{B}}_3$ are shown in Figure~\ref{fig:pointed3}. Points that satisfy
the same literals are given identical colours. Let us denote by $B$, $G$, and $W$ the set of literals true on the black, grey, and white
point, respectively. Let us suppose that Hercules has chosen the pointed model $(\frm B, B)$ shown on the right
of the doted line. The Hydra responds with the pointed models $(\frm A_B, B^{\rm i})$, $(\frm A_W,B^{\rm i})$, and $(\frm A_G,B^{\rm i})$
shown on the left.
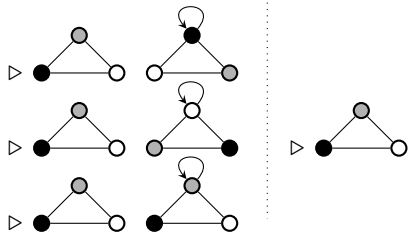
\begin{figure} [htbp]
\begin{center}
\begin{tikzpicture}
[
help/.style={circle,draw=white!100,fill=white!100,thick, inner sep=0pt,minimum size=1mm},
white/.style={circle,draw=black!100,fill=white!100,thick, inner sep=0pt,minimum size=2mm},
black/.style={circle,draw=black!100,fill=black!100,thick, inner sep=0pt,minimum size=2mm},
grey/.style={circle,draw=black!100,fill=black!30,thick, inner sep=0pt,minimum size=2mm},
]

\node at (1,1) [black](downnonreflLeft) [ label=180:$\triangleright$]{};
\node at (1.5,1.5) [grey](downnonreflMid) [ ]{};
\node at (2,1) [white](downnonreflRight) [ ]{};

\node at (2.5,1) [black](downreflLeft) [ ]{};
\node at (3,1.5) [grey](downreflMid) [ ]{};
\node at (3.5,1) [white](downreflRight) [ ]{};

\node at ( 4,4) [help](helpUpDotted) []{};
\node at ( 4,1) [help](helpDownDotted) []{};

\begin{scope}[>=stealth, auto]
\draw [-] (downnonreflLeft) to  (downnonreflMid);
\draw [-] (downnonreflLeft) to  (downnonreflRight);
\draw [-] (downnonreflMid) to  (downnonreflRight);

\draw [-] (downreflLeft) to  (downreflMid);
\draw [-] (downreflLeft) to  (downreflRight);
\draw [-] (downreflMid) to  (downreflRight);

\draw [->] (downreflMid) [out=50, in=130, loop]to  (downreflMid);

\draw [ dotted] (helpUpDotted) to  (helpDownDotted);
\end{scope}
\node at (1,2) [black](midnonreflLeft) [label=180:$\triangleright$ ]{};
\node at (1.5,2.5) [grey](midnonreflMid) [ ]{};
\node at (2,2) [white](midnonreflRight) [ ]{};

\node at (2.5,2) [grey](midreflLeft) [ ]{};
\node at (3,2.5) [white](midreflMid) [ ]{};
\node at (3.5,2) [black](midreflRight) [ ]{};

\begin{scope}[>=stealth, auto]
\draw [-] (midnonreflLeft) to  (midnonreflMid);
\draw [-] (midnonreflLeft) to  (midnonreflRight);
\draw [-] (midnonreflMid) to  (midnonreflRight);

\draw [-] (midreflLeft) to  (midreflMid);
\draw [-] (midreflLeft) to  (midreflRight);
\draw [-] (midreflMid) to  (midreflRight);

\draw [->] (midreflMid) [out=50, in=130, loop]to  (midreflMid);

\end{scope}

\node at (1,3) [black](upnonreflLeft) [label=180:$\triangleright$ ]{};
\node at (1.5,3.5) [grey](upnonreflMid) [ ]{};
\node at (2,3) [white](upnonreflRight) [ ]{};

\node at (2.5,3) [white](upreflLeft) [ ]{};
\node at (3,3.5) [black](upreflMid) [ ]{};
\node at (3.5,3) [grey](upreflRight) [ ]{};

\begin{scope}[>=stealth, auto]
\draw [-] (upnonreflLeft) to  (upnonreflMid);
\draw [-] (upnonreflLeft) to  (upnonreflRight);
\draw [-] (upnonreflMid) to  (upnonreflRight);

\draw [-] (upreflLeft) to  (upreflMid);
\draw [-] (upreflLeft) to  (upreflRight);
\draw [-] (upreflMid) to  (upreflRight);

\draw [->] (upreflMid) [out=50, in=130, loop]to  (upreflMid);

\end{scope}


\node at (4.75,2) [black](kLeft) [label=180:$\triangleright$ ]{};
\node at (5.25,2.5) [grey](kMid) [ ]{};
\node at (5.75,2) [white](kRight) [ ]{};

\begin{scope}[>=stealth, auto]
\draw [-] (kLeft) to  (kMid);
\draw [-] (kLeft) to  (kRight);
\draw [-] (kMid) to  (kRight);

\end{scope}
\end{tikzpicture}
\end{center}
\caption{The  sets $\cls{A}^3$ and $\cls{B}^3$.}
\label{fig:pointed3}
\end{figure}

\end{example}

\noindent{\sc formula size game on models:}
We consider the {\fsgm} starting with  $\cls{A}_n$
 on the left and $\cls{B}_n$
on the right.

\begin{definition}\label{def:specialPairs}
A special pair of pointed models is a pair $\langle(\frm A_S, E),(\frm{B}, E')\rangle$ such that $E=E'$. 
\end{definition}

\begin{proposition}\label{propPairLeaf}
For any game tree $T$ for a {\fsgm} and any node $\eta$ of $T$, if there is a special pair
$\langle(\frm A_S, E^x),(\frm{B}, E)\rangle$ with   $(\frm A_S, E^x)\in \lft(\eta)$ and
$(\frm{B}, E)\in \rgt(\eta)$, then
\begin{enumerate}

\item
 Hercules did not play a literal move at $\eta$;
\item
if $x = {\rm i}$ and Hercules did not play an  $\exists$-move at $\eta$, then,
 for at least one successor $\eta_1$  of $\eta$, there is a special pair $\langle(\frm A_S, U^{\rm i}),(\frm{B}, U)\rangle$ 
such that $(\frm A_S, U^{\rm i})\in \lft(\eta_1)$ and $(\frm{B}, U^{\rm i})\in \rgt(\eta_1)$.

\end{enumerate}

\end{proposition}

\proof The first item is obvious. For the second item we have to consider $\vee$-, $\wedge$-, $\pd$-, $\Box$-, and $\forall$-moves. 
 If Hercules played either an $\vee$- or an $\wedge$-move
at $\eta$ it is clear that the statement is true. If Hercules played a $\pd$-move,
since $E^{\rm i})$ is a point in the irreflexive component of ${\Kpair{{\frm K}}}_n$,  he must have picked a successor $(\frm A_S, U^{\rm i})$
of $(\frm A_S, E^{\rm i})$ with $ U \not= E$.
Since the Hydra plays greedily, we know that she is going to pick, among others, the pointed model $(\frm{B}, U)\in \Box(\frm{B}, E)$ and the statement
follows. The cases for $\Box$- and $\forall$-moves are treated similarly. 
\endproof

\begin{lemma}\label{lemmaExistsMoves}
For any classes of pointed models  $\cls L$ and $\cls R$ such that $\cls{A}_n\subseteq \cls{L}$, $\cls{B}_n\subseteq \cls{R}$, and Hercules has a winning strategy in the $( \languni,\langle \cls{L},\cls{R}\rangle)$-{\fsgm}, if
 $T$ is a closed game tree for this game and the Hydra played greedily, then $T$ has at least one node that is an 
$\exists$-move.
\end{lemma} 

\proof
Let $\rho$ denote the root of  $T$ and let  us fix a special pair  $\langle(\frm A_S, E^x),(\frm{B}, E)\rangle$ with $(\frm A_S, E^x)\in \lft(\rho)$ and 
$(\frm{B}, E)\in \rgt(\rho)$.
If we assume that Hercules did not play an $\exists$-move during the game, then
we see, using Proposition~\ref{propPairLeaf}, that $T$ is not a closed game tree, which is a contradiction.
\endproof

\begin{lemma}\label{lemmaLowerBound}
For any classes of pointed models  $\cls L$ and $\cls R$ such that $\cls{A}_n\subseteq \cls{L}$, $\cls{B}_n\subseteq \cls{R}$, and Hercules has a winning strategy in the $( \languni,\langle \cls{L},\cls{R}\rangle)$-{\fsgm}, if
 $T$ is a closed game tree for this game and the Hydra played greedily, then $T$ has at least $n$ nodes.
\end{lemma} 
The proof of the lemma revolves around the notion of {\em weight function}---a popular tool in Boolean function complexity~\cite{jukna} where it is
often called {\em complexity functional}. Intuitively, a weight function
is a tool that allows us to formulate proofs by induction on a notion of  ``progress during a \fsgm''.
\begin{definition}\label{def:weightf} For any finite binary tree $T$, a weight function $f$ for $T$ is a function
that assigns to any node $\eta$  of $T$  a non-negative real number such that 
\begin{enumerate}
\item\label{itWeightOne} if $\eta$ is a leaf, then $f(\eta)\leq 1$;
\item\label{itWeightTwo} if $\eta$ has two immediate successors $\eta_1$ and $\eta_2$, then $f(\eta)\leq f(\eta_1)+f(\eta_2)+1$;
\item\label{itWeightThree} if $\eta$ has one immediate successor $\eta_1$, then $f(\eta)\leq f(\eta_1)+1$.
\end{enumerate} 
\end{definition}
\begin{lemma}\label{lemma:weightf}
For any finite binary tree $T$ and any weight function $f$ for $T$,
if $\rho$ is the root of $T$, then $T$ has at least $f(\rho)$ nodes.
\end{lemma}
\proof
An easy induction on the number of nodes in $T$.
\endproof

In order to prove Lemma~\ref{lemmaLowerBound} we  define a suitable weight function on the nodes of $T$ as follows.
For a node $\eta$ of $T$, we let $f(\eta)$ be the number of $S \in W_n$ such that there is at least one special pair $\langle(\frm A_S, E^x),(\frm{B}, E)\rangle$ with $(\frm A_S, E^x)\in\lft({\eta})$ and $(\frm{B}, E)\in\rgt({\eta})$.

Obviously, if $\rho$ denotes the 
root of $T$, then $f(\rho)=n$.
So, it remains to check the following.

\begin{lemma}\label{lemmIsFunct}
The function $f$ defined above is a complexity functional.
\end{lemma}

\proof
We need to show that $f$ satisfies the three items 
from Definition~\ref{def:weightf}.\\

\noindent\eqref{itWeightOne} If $\eta$ is a leaf, then it is immediate from the first item of Proposition~\ref{propPairLeaf}  that there is no special pair 
$\langle(\frm A_S, E^x),(\frm{B}, E)\rangle$ with $(\frm A_S, E^x)\in\lft({\eta})$ and $(\frm{B}, E)\in\rgt({\eta})$.
Hence, $f(\eta)\leq 1$.\\

\noindent\eqref{itWeightTwo} If $\eta$ has two immediate successors $\eta_1$ and $\eta_2$, then $\eta$ represents either an $\vee$- or an $\wedge$-move.
It is easily seen that $f(\eta_1) + f(\eta_2)\geq f(\eta)$ and, therefore, the second condition of Definition~\ref{def:weightf}
is fulfilled.\\

\noindent\eqref{itWeightThree} If $\eta$ has one immediate successor $\eta'$, then $\eta$ represents a $\forall$-, $\exists$-, $\Box$-, or 
$\pd$-move.
Let $\Gamma $ be the set of all $S \in W_n$ such that there is a special pair $\langle(\frm A_S, E^x),(\frm{B}, E)\rangle$
 with
 $(\frm A_S, E^x)\in\lft({\eta})$ and $(\frm{B}, E)\in\rgt({\eta})$, and define $\Gamma '$ analogously with $\eta'$ in place of $\eta$.
In each case, we claim that there is $\Delta \subseteq \Gamma$ with $|\Delta |\leq 1$ such that $\Gamma \setminus \Delta \subseteq \Gamma'$, from which we obtain $f(\eta')\geq f(\eta) + 1$.
We consider the following cases.
\\

\noindent {\sc $\eta$ is a $\forall$-move.} Let $S\in \Gamma$, so that for some $E$ we have that  $(\frm A_S, E^x)\in\lft({\eta})$ and $(\frm{B}, E)\in\rgt({\eta})$.
If Hercules picks $(\frm{B}, U)$ as a successor of $(\frm{B}, E)$, then, since the Hydra is playing greedily,
we know that the pointed model $(\frm A_S, U^x )$, where $U$ is a point in the non-reflexive component of $\frm A_S$, is going to be in  $\lft({\eta}')$. Therefore, $U$ witnesses that $S\in \Gamma'$, and $\Gamma \subseteq \Gamma'$.\\

\noindent {\sc $\eta$ is an $\exists$-move.} Let us consider a special pair
$\langle(\frm A_S, E^x),(\frm{B}, E)\rangle$ with
 $(\frm A_S, E^x)\in\lft({\eta})$ and $(\frm{B}, E )\in\rgt({\eta})$.
If Hercules places $(\frm A_S, U^y )$ in $\lft({\eta}')$ as the successor of $(\frm A_S, E^x )$, then the Hydra's 
greedy strategy guarantees that $(\frm{B}, U )\in\rgt(\eta')$, and $\Gamma \subseteq \Gamma'$.\\

\noindent {\sc $\eta$ is a $\Box$-move.} Let $S\in \Gamma$, so that $(\frm A_S, E^x )\in\lft({\eta})$ and $(\frm{B}, E )\in\rgt({\eta})$ for some $E$.
If Hercules picks $(\frm{B}, U )$ as a successor of $(\frm{B}, E )$, then, the Hydra, by playing greedily,
is going to place a pointed model $(\frm A_S, U^x)$ in  $\lft({\eta}')$ because there are $U$-points in both the
reflexive and the non-reflexive component of $\frm A_S$. Therefore, $\Gamma \subseteq \Gamma'$.\\

\noindent {\sc $\eta$ is a $\pd$-move.}
Partition $\Gamma$ into two subsets $\Sigma $ and $\Delta = \Gamma \setminus \Sigma$, where $S\in \Sigma$ if there are $E\not=F$ and $x,y$ such that $(A_S,E^x) \in \lft(\eta)$, $(\frm B,E) \in \rgt(\eta)$, and $(A_S,F^y) \in \lft(\eta')$. For such an $S$, the Hydra's greedy strategy implies that $(\frm{B}, F)$ is also going to be among the 
successors of $(\frm{B}, E)$ picked by her and thus $S\in \Gamma'$. Since $S\in\Sigma$ was arbitrary, $\Sigma\subseteq \Gamma'$.

If $|\Delta| \leq 1$ we are done, since then $\Gamma \setminus \Delta \subseteq \Gamma'$.
So assume otherwise, and let $S\in \Delta$.
Since $S\in \Gamma$ there must be a set of variables $E$ and $x\in \{{\rm i},{\rm r}\}$ such that $(\frm A_S,E^x) \in \lft  ( \eta ) $ and $(\frm B, E) \in \rgt (\eta)$.
Let $(\frm A_S,F^x) \in \lft (\eta')$ be the successor chosen by Hercules; since $S \not \in \Sigma$ we must have $F=E$, hence $E^x$ is the unique reflexive point of $\frm A_S$ so that $E^x = S^{\rm r}$.

Using the assumption that $|\Delta| > 1$, let $U \not= S$ be another element of $\Delta$.
As above we have that $(\frm B, U) \in \rgt ( \eta ) $, hence the Hydra's greedy strategy implies that $(\frm B, S) \in \rgt ( \eta' ) $. Since also $(\frm A_S,S^{\rm r}) \in \lft (\eta')$, we have that $S \in \Gamma'$, as needed.
\endproof

Lemma~\ref{lemmaLowerBound} is an immediate consequence of Lemmas \ref{lemma:weightf} and \ref{lemmIsFunct}. With this, we have established all claims of Theorem~\ref{thm:noncol}.

\section{The transfer axioms}\label{secGenTrans}

In this section we consider what we call the {\em transfer axioms,} defi\-ned as ${\rm TA}(m,n) = \pd^m p \to \pd ^n p$, where $m \not = n\in\mathbb N$; since we treat $\varphi \to \psi$ as an abbreviation, we can rewrite these axioms as $\Box^m \overline p \vee \pd^n p$.
It is well-known that ${\rm TA}(m,n)$ defines the first-order property of {\em $(m,n)$-transfer}~\eqref{eqGT} from the introduction.
As special cases we have that $(2,1)$-transfer is just transi\-ti\-vi\-ty and $(0,1)$-transfer is reflexivity. Instead of $(m,n)$-transfer we write {\em $n$-reflexivity} when $m=0$, {\em $m$-recurrence} when $n=0$, {\em $(m,n)$-transitivity} when $m>n>0$ and {\em $(m,n)$-density} when $0<m<n$.

Our goal is to prove the following.

\begin{theorem}\label{theoTransfer} For any $n \not = m \in\mathbb N$, $\Box^m \overline p \vee \pd^n p$ is absolutely minimal among all formulas defining $(m,n)$-transfer.
\end{theorem}

The proof that for each $m,n\geq0$, $\pd^m p\rightarrow\pd^n p$ is essentially the shortest formula defining $(m,n)$-transfer is split in four parts according to the ordering between $m$ and $n$.

\subsection{Generalized density axioms}\label{secDens}

First we consider the generalized density axioms, i.e.~$(m,n)$-transfer when $0<m<n$.
We prove that Theorem \ref{theoTransfer} holds in this case by considering a suitable formula-complexity game.
Specifically, Hercules and the  Hydra play a $( \lang_{\pd},\langle \cls{A},\cls{B}\rangle)$-{\fsgf} where
$\cls{A}=\{\frm{A}_1, \ldots \frm{A}_{m+1}\}$ and $\cls{B}$ contains a single element $\frm{B}$.
These frames are shown in the left rectangle in Figure~\ref{fig:nbiggerthanm} and separated by the dotted line.
\begin{figure} 
\begin{center}
\begin{tikzpicture}
[
help/.style={circle,draw=white!100,fill=white!100,thick, inner sep=0pt,minimum size=1mm},
white/.style={circle,draw=black!100,fill=white!100,thick, inner sep=0pt,minimum size=1.5mm},
black/.style={circle,draw=black!100,fill=black!100,thick, inner sep=0pt,minimum size=1.5mm},
xscale=0.84,
yscale=0.84
]


\node at (.5,1.5) [help](a1) [label=270: $\frm{A}_1$]{};
\node at (.5,2) [white](roota1) []{};
\node at (0,2) [white](loopa1rootl) [ ]{};

\node at (.5,2.5) [white](a11) []{};
\node at (1,2.5) [white](loopa11r) []{};

\node at (.5,3.5) [white](a12) []{};
\node at (1,4) [white](loopa12r) []{};

\node at (.5,4) [white](a13) []{};
\node at (0,4) [white](loopa13) [ ]{};

\begin{scope}[>=stealth, auto]
\draw [->] (roota1) to  (loopa1rootl);
\draw [->] (loopa1rootl) [out=230, in=310, loop]to  (loopa1rootl);
\draw [->] (roota1) to  (loopa11r);
\draw [->] (loopa11r) [out=320, in=40, loop]to  (loopa1rootl);

\draw [->] (roota1) to  (a11);

\draw [-,dotted,thick] (a11) to  (a12);
\draw [-,dotted,thick] (loopa11r) to  (loopa12r);

\draw [->] (a12) to  (a13);
\draw [->] (loopa12r) [out=50, in=130, loop]to  (loopa12r);

\draw [->] (loopa12r) to  (a13);
\draw [->] (a13) to  (loopa13);

\draw [->] (loopa13) [out=50, in=130, loop]to  (loopa13);

\end{scope}
\node at (2,1.5) [help](a2) [label=270: $\frm{A}_2$]{};
\node at (2,2) [white](roota2) []{};
\node at (1.5,2) [white](loopa2rootl) [ ]{};

\begin{scope}[>=stealth, auto]
\draw [->] (roota2) to  (loopa2rootl);
\draw [->] (loopa2rootl) [out=230, in=310, loop]to  (loopa2rootl);
\end{scope}

\node at (3,1.5) [help](a3) [label=270: $\frm{A}_3$]{};
\node at (3,2) [white](roota3) []{};
\node at (2.5,2) [white](loopa3rootl) [ ]{};
\node at (3,2.5) [white](a31) []{};

\begin{scope}[>=stealth, auto]
\draw [->] (roota3) to  (loopa3rootl);
\draw [->] (roota3) to  (a31);
\draw [->] (loopa3rootl) [out=230, in=310, loop]to  (loopa3rootl);
\end{scope}

\node at (3.5,2) [help](ldots) []{$\ldots$};

\node at (4.5,1.5) [help](am1) [label=270: $\frm{A}_{m+1}$]{};
\node at (4.5,2) [white](rootam1) []{};
\node at (4,2) [white](loopam1rootl) [ ]{};
\node at (4.5,2.5) [white](am11) []{};
\node at (4.5,3) [white](am12) []{};
\node at (4.5,3.5) [white](am13) []{};
\begin{scope}[>=stealth, auto]
\draw [->] (rootam1) to  (loopam1rootl);
\draw [->] (rootam1) to  (am11);
\draw [->] (loopam1rootl) [out=230, in=310, loop]to  (loopam1rootl);
\draw [-, dotted, thick] (am11) to  (am12);
\draw [->] (am12) to  (am13);
\end{scope}


\node at ( 5,4) [help](helpUpDotted) []{};
\node at ( 5,1.15) [help](helpDownDotted) []{};

\begin{scope}[>=stealth, auto]



\draw [ dotted] (helpUpDotted) to  (helpDownDotted);


\end{scope}

RIGHT FRAME

\node at (5.8,1.5) [help](B) [label=270: $\frm{B}$]{};
\node at (5.8,2) [white](rootb) []{};
\node at (5.3,2) [white](loopbrootl) [ ]{};

\node at (5.8,2.5) [white](b1) []{};

\node at (5.8,3.5) [white](b2) []{};

\node at (5.8,4) [white](b3) []{};
\node at (5.3,4) [white](loopb3) [ ]{};

\begin{scope}[>=stealth, auto]
\draw [->] (rootb) to  (loopbrootl);
\draw [->] (loopbrootl) [out=230, in=310, loop]to  (loopbrootl);

\draw [->] (rootb) to  (b1);

\draw [-,dotted,thick] (b1) to  (b2);

\draw [->] (b2) to  (b3);

\draw [->] (b3) to  (loopb3);

\draw [->] (loopb3) [out=50, in=130, loop]to  (loopb3);

\end{scope}

\node at (6.1,2)[help](Clengthdown1)[]{};
\node at (6.3,2)[help](Clengthdown2)[]{};
\node at (6.5,2)[help](Clengthdown3)[]{};

\node at (6.1,4)[help](Clengthup1)[]{};
\node at (6.3,4)[help](Clengthup2)[]{};
\node at (6.5,4)[help](Clengthup3)[]{};
\begin{scope}[>=stealth, auto]
\draw [<->, dashed] (Clengthdown2) to node[swap]{$m$} (Clengthup2);
\draw[-](Clengthdown1) to (Clengthdown3);
\draw[-](Clengthup1) to (Clengthup3);
\end{scope}

\node at ( 6.9,.9) [help](mhelpDown) []{};
\node at ( 6.9,4.5) [help](mhelpUp) []{};
\node at ( -0.25,4.5) [help](mhelpleftUp) []{};
\node at (- 0.25,.9) [help](mhelpleftDown) []{};
\node at ( 14.1,4.5) [help](mhelpRightUp) []{};
\node at ( 14.1,.9) [help](mhelpRightDown) []{};

\begin{scope}[>=stealth, auto]

\draw  (mhelpUp) to  (mhelpDown);
\draw  (mhelpleftUp) to  (mhelpleftDown);

\draw  (mhelpRightUp) to  (mhelpRightDown);

\draw  (mhelpUp) to  (mhelpRightUp);
\draw  (mhelpUp) to  (mhelpleftUp);
\draw   (mhelpDown) to (mhelpleftDown);
\draw   (mhelpDown) to (mhelpRightDown);
\end{scope}


\node at (7.7,1.6) [help](ma1) [label=270:$\mods{\frm A}_1$]{};
\node at (7.7,2) [white](mroota1) [label=0:$\triangleleft$]{};
\node at (7.2,2) [black](mloopa1rootl) [ ]{};

\node at (7.7,2.5) [white](ma11) []{};
\node at (8.2,2.5) [black](mloopa11r) []{};

\node at (7.7,3.5) [white](ma12) []{};
\node at (8.2,4) [black](mloopa12r) []{};

\node at (7.7,4) [white](ma13) []{};
\node at (7.2,4) [white](mloopa13) [ ]{};

\begin{scope}[>=stealth, auto]
\draw [->] (mroota1) to  (mloopa1rootl);
\draw [->] (mloopa1rootl) [out=230, in=310, loop]to  (mloopa1rootl);
\draw [->] (mroota1) to  (mloopa11r);
\draw [->] (mloopa11r) [out=320, in=40, loop]to  (mloopa1rootl);

\draw [->] (mroota1) to  (ma11);

\draw [-,dotted,thick] (ma11) to  (ma12);
\draw [-,dotted,thick] (mloopa11r) to  (mloopa12r);

\draw [->] (ma12) to  (ma13);
\draw [->] (mloopa12r) [out=50, in=130, loop]to  (mloopa12r);

\draw [->] (mloopa12r) to  (ma13);
\draw [->] (ma13) to  (mloopa13);

\draw [->] (mloopa13) [out=50, in=130, loop]to  (mloopa13);

\end{scope}
\node at (9.2,1.6) [help](ma2) [label=270: $\mods{\frm A}_2$]{};
\node at (9.2,2) [white](mroota2) [label=0:$\triangleleft$]{};
\node at (8.7,2) [black](mloopa2rootl) [ ]{};

\begin{scope}[>=stealth, auto]
\draw [->] (mroota2) to  (mloopa2rootl);
\draw [->] (mloopa2rootl) [out=230, in=310, loop]to  (mloopa2rootl);
\end{scope}

\node at (10.5,1.6) [help](ma3) [label=270:$\mods{\frm A}_3$]{};
\node at (10.5,2) [white](mroota3) [label=0:$\triangleleft$]{};
\node at (10,2) [black](mloopa3rootl) [ ]{};
\node at (10.5,2.5) [white](ma31) []{};

\begin{scope}[>=stealth, auto]
\draw [->] (mroota3) to  (mloopa3rootl);
\draw [->] (mroota3) to  (ma31);
\draw [->] (mloopa3rootl) [out=230, in=310, loop]to  (mloopa3rootl);
\end{scope}

\node at (11.25,2) [help](mldots) []{$\ldots$};

\node at (12.15,1.6) [help](mam1) [label=270:$\mods{\frm A}_{m+1}$]{};
\node at (12.15,2) [white](mrootam1) [label=0:$\triangleleft$]{};
\node at (11.65,2) [black](mloopam1rootl) [ ]{};
\node at (12.15,2.5) [white](mam11) []{};
\node at (12.15,3) [white](mam12) []{};
\node at (12.15,3.5) [white](mam13) []{};
\begin{scope}[>=stealth, auto]
\draw [->] (mrootam1) to  (mloopam1rootl);
\draw [->] (mrootam1) to  (mam11);
\draw [->] (mloopam1rootl) [out=230, in=310, loop]to  (mloopam1rootl);
\draw [-, dotted, thick] (mam11) to  (mam12);
\draw [->] (mam12) to  (mam13);
\end{scope}


\node at ( 12.7,4) [help](mhelpUpDotted) []{};
\node at ( 12.7,1) [help](mhelpDownDotted) []{};

\begin{scope}[>=stealth, auto]



\draw [ dotted] (mhelpUpDotted) to  (mhelpDownDotted);


\end{scope}

RIGHT MODEL

\node at (13.5,1.6) [help](mB) [label=270: $\mods{\frm B}$]{};
\node at (13.5,2) [white](mrootb) [label=0:$\triangleleft$]{};
\node at (13,2) [black](mloopbrootl) [ ]{};

\node at (13.5,2.5) [white](mb1) []{};

\node at (13.5,3.5) [white](mb2) []{};

\node at (13.5,4) [white](mb3) []{};
\node at (13,4) [white](mloopb3) [ ]{};

\begin{scope}[>=stealth, auto]
\draw [->] (mrootb) to  (mloopbrootl);
\draw [->] (mloopbrootl) [out=230, in=310, loop]to  (mloopbrootl);

\draw [->] (mrootb) to  (mb1);

\draw [-,dotted,thick] (mb1) to  (mb2);

\draw [->] (mb2) to  (mb3);

\draw [->] (mb3) to  (mloopb3);

\draw [->] (mloopb3) [out=50, in=130, loop]to  (mloopb3);

\end{scope}



\end{tikzpicture}
\end{center}
\caption{The frames ${\frm A}_1$, $\ldots$, ${\frm A}_{m+1}$ and ${\frm B}$ and the pointed models based on them. }
\label{fig:nbiggerthanm}
\end{figure}
$\frm{A}_{1}$ is constructed so that the vertical path leading from the lowest non-reflexive point to
the uppermost non-reflexive one consists of $m$ steps whereas the rightmost path that starts and ends
respectively with these two points consists of $n$ steps (not counting the reflexive steps) and every point on this rightmost path is reflexive. 
The frame $\frm{B}$ is obtained from $\frm{A}_1$ by simply erasing the latter path.
 Each $\frm{A}_i$, for $2\leq i \leq m+1$, contains a vertical path 
of  $i-2$ steps. Obviously, $\pd^m p\rightarrow \pd^n p$ is valid in all frames in $\cls{A}$  and not valid on $\frm{B}$.\\

\noindent{\sc selection of the models on the right:}
If Hercules wishes to win the game, he must choose his pointed models with some care.

\begin{lemma}\label{lemA1Bdens}
In any winning strategy for Hercules for an $( \lang_{\pd},\langle \cls{L},\cls{R}\rangle)$-{\fsgf} in which $\frm{A}_1\in \cls{L}$ and $\frm{B}\in \cls{R}$, Hercules must pick a pointed model $(\mods{\frm B},\point)$ based on the lowest irreflexive point in $\frm{B}$.
\end{lemma}

\proof
It is easy to see that Hercules is not going to select a pointed model that is not based on the lowest non-reflexive point in $\frm{B}$ because the Hydra can always reply with a bisimilar pointed model based on $\frm{A}_1$. 
\endproof

\noindent{\sc selection of models on the left:}
The Hydra replies with the pointed models shown on the left of the dotted line in the right rectangle in Figure~\ref{fig:nbiggerthanm}. 
 She has constructed them as follows. Using the fact that $\frm{B}$ is a sub-structure of $\frm{A}_1$, the Hydra
makes sure that the same points in $\mods{\frm{A}}_1$ and $\mods{\frm{B}}$ satisfy the same literals; moreover, the black points in both models
satisfy the same literals, too. The models $\mods{\frm{A}}_i$ for $2\leq i\leq m+1$ receive valuations that make them initial segments of the vertical path in
$\mods{\frm{B}}$, i.e., the lowest  non-reflexive point in any $\mods{\frm{A}}_i$ and the lowest non-reflexive point in $\mods{\frm{B}}$ satisfy the same literals and similarly
for their vertical successors.
When the Hydra chooses her pointed models in this way, we say she {\em mimics} Hercules' choice.
\medskip

\noindent{\sc formula size game on models:}
We consider the {\fsgm} starting with  
$(\mods{\frm A}_1, \triangleright),\ldots, (\mods{\frm A}_{m+1}, \triangleright)$ on the left and $(\mods{\frm B}, \triangleright)$
on the right.
First we show that there are some constraints on the moves that Hercules may make.

\begin{lemma}\label{lemNoLeft}
Let $\cls L$, $\cls R$ be classes of models such that Hercules has a winning strategy for the $(\lang_\pd, \langle \cls L,\cls R \rangle )$-{\fsgm}.
Let $T$ be any closed game tree on which the Hydra played greedily and $\eta$ be any position of $T$ such that  
 $(\mods{\frm{B}}, \triangleright) \in \rgt(\eta)$ while 
$(\frm{A}_i, \triangleright) \in \lft(\eta)$ for some $i$ with $1 \leq i\leq m+1$.
\begin{enumerate}

\item\label{itNoLeftOne} If Hercules played a $\pd$-move at $\eta $ then he did not pick the left lowest reflexive point in $\mods{\frm{A}}_i$, and if $i = 1$ then he picked the bottom-right reflexive point on $\mods{\frm A}_1$.

\item\label{itNoLeftTwo} If Hercules played a $\Box$-move at $\eta $ then he did not pick the left lowest reflexive point in $\mods{\frm{B}}$.

\end{enumerate}

\end{lemma}

\proof
If Hercules  picks the left lowest reflexive point when playing such a move, the Hydra is going to reply with the same point in $\mods{\frm{B}}_1$ and obtain bisimilar pointed models on each side.
If $i = 1$ and Hercules picks the unique irreflexive successor on $\mods{\frm A}_1$, then Hydra can reply with the irreflexive successor on $\mods{\frm B}$, which means by Corollary~\ref{lemHercLose} that Hercules cannot win. The second claim is symmetric.
\endproof

\begin{lemma}\label{lemmDensNoModal}
Suppose that $\cls L$, $\cls R$ are classes of models and Hercules has a winning strategy for the $(\lang_\pd, \langle \cls L,\cls R \rangle )$-{\fsgm}.
If $T$ is any closed game tree in which the Hydra played greedily and $\eta$ is any position of $T$ such that $(\mods{\frm{B}},\triangleright)\in \rgt(\eta)$, then
\begin{enumerate}

\item\label{itDensNoModalOne}
if $(\mods {\frm{A}}_1, \triangleright)\in \lft (\eta)$, then 
Hercules did not play a $\Box$-move on $\eta$;

\item \label{DensNoModalTwo}
if $(\mods {\frm{A}}_2,\triangleright) \in \lft(\eta)$, 
then Hercules did not play a $\pd$-move on $\eta$.

\end{enumerate}
\end{lemma}

\proof
The first claim is immediate from the fact that  if Hercules played a $\Box$-move, the Hydra can reply with 
the same point in $\mods{\frm{A}}_1$ and obtain bisimilar pointed models on each side. For the second, Hercules is forced to pick the reflexive point in $\mods{\frm{A}}_2$ when playing a $\pd$-move which contradicts Lemma \ref{lemNoLeft}.
\endproof

With this we can establish lower bounds on the number of moves of each type that Hercules must make, as established by the proposition below.

\begin{proposition}\label{propDensBounds}
Let $\cls L$, $\cls R$ be classes of models such that Hercules has a winning strategy for the $(\lang_\pd, \langle \cls L,\cls R \rangle )$-{\fsgm} and let $T$ be a closed game tree in which the Hydra played greedily.
\begin{enumerate}

\item\label{itDensDisj} If $\{(\mods{\frm{A}}_1, \triangleright), (\mods{\frm{A}}_2,\triangleright)\}\subseteq \cls{L}$ and $(\frm{B}, \triangleright)\in \cls{R}$, then Hercules made
at least one $\vee$-move during the game.

\item \label{itDensDiam} If $(\mods {\frm{A}}_1, \triangleright)\in \cls L$, and $(\mods{\frm{B}},\triangleright)\in \cls R$, then $T$ has modal depth at least $n$, at least $n$ $\pd$-moves and one literal.

\item \label{itDensBox} If $\{(\mods{\frm{A}}_2, \triangleright),\ldots, (\mods{\frm{A}}_{m+1},\triangleright)\}\subseteq \cls{L}$ and $(\mods{\frm B},\point) \in \cls R$, then Hercules made
at least $m$ $\Box$-moves during the game.

\end{enumerate}

\end{proposition}

\proof\

\noindent \eqref{itDensDisj}
By Lemma \ref{lemmDensNoModal}, Hercules cannot play a modality as long as 
$(\mods{\frm{A}}_1, \triangleright), (\mods{\frm{A}}_2,\triangleright)$ 
are both on the left and $ (\frm{B}, \triangleright)$ on the right, and  the three satisfy the same literals, so that he cannot play a literal either.
Playing a $\wedge$-move would lead to at least one new game position that is the same as the previous one.
Hence, every winning strategy for Hercules must `separate' $(\frm{A}_1, \triangleright)$, from $(\frm{A}_2,\triangleright)$
with an $\vee$-move.
\smallskip

\noindent \eqref{itDensDiam} Note that $(\mods{\frm{A}}_1,\triangleright)$ and $(\mods{\frm{B}},\triangleright)$ satisfy the 
same literals and $\vee$- and $\wedge$-moves lead to at least one new game-position  in which   $(\mods{\frm{A}}_1,\triangleright)$
is on the left and $(\mods{\frm{B}},\triangleright)$ is on the right.
By Lemma~\ref{lemmDensNoModal}.\ref{itDensNoModalOne}, Hercules cannot play a $\Box$-move in any of these positions.
Thus Hercules must perform a $\pd$-move in a position in which  $(\mods{\frm{A}}_1,\triangleright)$
is on the left and $(\mods{\frm{B}},\triangleright)$ is on the right. By Lemma \ref{lemNoLeft}.\ref{itNoLeftOne} he is going to pick the first reflexive point on the rightmost path in $\mods{\frm{A}}_1$.

The Hydra replies with, among others, the left lowest reflexive point in $\mods{\frm{B}}$. Since this point satisfies the same literals
as the reflexive points lying on the rightmost path in $\mods{\frm{A}}_1$, Hercules cannot play a literal-move; moreover,
$\vee$-, $\wedge$- and $\Box$-moves lead to at least one new game position that is essentially the same as the previous one.
In the case of $\Box$-moves this is true because, when playing such a move, Hercules must stay in the lowest reflexive point
in $\mods{\frm{B}}$ while the Hydra can stay in the current reflexive point on the rightmost path in $\mods{\frm{A}}_1$. Hence, he must make at least $n-1$ subsequent $\pd$-moves
to reach a point in $\mods{\frm{A}}_1$ that differs on a literal from the lowest reflexive point in $\mods{\frm{B}}$. Finally he must play a literal, as no other move can close the tree.\smallskip

\noindent \eqref{itDensBox}
Fix $i\in [2,m+1]$. Let $w_1,\hdots, w_{i-1}$ enumerate the vertical path of $\frm A_i$ starting at the root, and similarly let $v_1,\ldots,v_m$ enumerate the vertical path of $\frm B$. Let $\pointed w_j = (\mods{\frm A}_i,w_j )$ and $\pointed v_j = (\mods{\frm B},v_j )$.

Say that a branch $\overrightarrow \nu = (\nu_0, \ldots ,\nu _k)$ on $T$ is {\em $i$-critical} if there exists $j\in [1,i)$ with $\pointed w_j \in \lft(\nu _k)$, $\pointed v_j \in \rgt(\nu _k)$ and Hercules has played exactly $j-1$ modal moves on $ \nu_1,\ldots,\nu_{k-1}$. Since $T$ is finite and the singleton branch consisting of the root is $i$-critical, we can pick a maximal $i$-critical branch $\overrightarrow \eta = (\eta_0, \ldots ,\eta_\ell)$ for some value of $j$.

We claim that $j=i-1$ and Hercules plays a $\Box$-move on $\eta_\ell$. Since $T$ is closed $\eta_\ell$ cannot be a head, but $\pointed w_j$ and $\pointed v_j$ share the same valuation so it cannot be a stub either, thus $\eta_\ell$ is not a leaf. If Hercules played an $\wedge$- or an $\vee$-move then $\eta_\ell$ would have a daughter giving us a longer $i$-critical branch. Thus Hercules played a modality on $\eta_\ell$. If $j<i-1$ then for the unique daughter $\eta'$ of $\eta_\ell$ we have that $\pointed w_{j+1} \in \lft(\eta')$ and $\pointed v_{j+1} \in \rgt(\eta')$, where in the case of $j = 0$ we use Lemma \ref{lemNoLeft} and otherwise there simply are no other options for Hercules; but this once again gives us a longer $i$-critical branch. Thus $j=i-1$; but then Hercules is not allowed to play $\pd$, as there is a pointed model on the left without successors, so he played a $\Box$-move on $\eta_\ell$.

We conclude that for each $i\in [2,m+1]$ there is an instance of $\Box$ with modal depth exactly $i-1$, which implies that each instance is distinct.
\endproof

\noindent With this we prove Theorem \ref{theoTransfer} in the case $0<m<n$.

\proof
If $0<m<n$ we consider the $(\lang_\pd,\langle \cls A,\cls B\rangle)$-{\fsgf} with $\cls A$, $\cls B$ as depicted in Figure \ref{fig:nbiggerthanm}. By Lemma \ref{lemA1Bdens} Hercules chooses some pointed model $\mods{\frm B}$ based on the irreflexive point at the bottom of $\frm B$, and Hydra replies by mimicking Hercules' pointed models. Then by Proposition \ref{propDensBounds} Hercules must play at least one disjunction, one literal, $n$ $\pd$-moves, modal depth at least $n$, and $m$ $\Box$-moves. By Theorem \ref{thrm: validityGames}, any formula valid on every frame of $\cls A$ and no frame of $\cls B$ must satisfy these bounds; but the frames in $\cls A$ satisfy the $(m,n)$-transfer property while those in $\cls B$ do not.
\endproof

\subsection{Generalized transitivity axioms}\label{sect:trans}

Next we treat Theorem \ref{theoTransfer} in the case where $0<n<m$.
As before, we do so by considering a suitable $( \lang_{\pd},\langle \cls{A},\cls{B}\rangle)$-{\fsgf} where
$\cls{A}=\{\frm{A}_1, \ldots \frm{A}_{m+1}\}$ and $\cls{B}$ contains a single element $\frm{B}$, but now using the frames shown in Figure~\ref{fig:nsmallerthanm}.
The frame $\frm{A}_1$ is based on a right-angled triangle in which  the sum of the relation steps in the legs is $m$ whereas the number of relation
steps in the hypotenuse is $n$; moreover, each path on the left of the hypotenuse that shares nodes with it  consist of $n$ relation steps, too.
The frame $\frm{B}$ is obtained from $\frm{A}_1$ by ``separating'' the hypotenuse from the horizontal leg and erasing the points that do not lie either
on the hypotenuse or on the legs of $\frm{A}_1$. Each $\frm{A}_i$, for $2\leq i \leq m+1$, contains a vertical path 
of  $i-2$ relation steps and a diagonal one of $n$ relation steps. 
Obviously, $\pd^m p\rightarrow \pd^n p$ is valid in all frames in $\cls{A}$  and not valid on $\frm{B}$.
\begin{figure}[h] 
\begin{center}
\begin{tikzpicture}
[
help/.style={circle,draw=white!100,fill=white!100,thick, inner sep=0pt,minimum size=1mm},
white/.style={circle,draw=black!100,fill=white!100,thick, inner sep=0pt,minimum size=2mm},
black/.style={circle,draw=black!100,fill=black!100,thick, inner sep=0pt,minimum size=2mm},
rect/.style={rectangle,draw=black!100,fill=black!20,thick, inner sep=0pt,minimum size=2mm},
xscale=0.85,
yscale=0.85
]


\node at (2,0) [help](A1) [label=90: $\frm{A}_1$]{};
\node at (2,1) [white](rootA1) [ ]{};

\node at (1.5,1.5) [white](rootA11) [ ]{};
\node at (1,2) [white](rootA111) [ ]{};

\node at (.5,2.5) [white](rootA1111) [ ]{};

\node at (2,2) [white](A11) [ ]{};

\node at (2,3) [white](A13) [ ]{};

\node at (2,3.75) [white](A14) [ ]{};

\node at (1.6,2) [white](A11r) [ ]{};
\node at (1.2,2.5) [white](A111r) [ ]{};
\node at (.75,3) [white](A1111r) [ ]{};
\node at (.25,3.5) [white](A11111r) [ ]{};

\node at (1.3,3) [white](A13r) [ ]{};
\node at (.6,3.7) [white](A133r) [ ]{};
\node at (1,3.75) [white](A14r) [ ]{};

\begin{scope}[>=stealth, auto]
\draw [->] (rootA1) to  (A11);
\draw [->] (rootA1) to  (A11r);
\draw [->] (rootA1) to  (rootA11);
\draw [-, dotted, thick] (rootA11) to  (rootA111);
\draw [->] (rootA111) to  (rootA1111);

\draw [->] (A11r) to  (A111r);
\draw [-, dotted, thick] (A11r) to  (A13r);
\draw [-, dotted, thick] (A11) to  (A13);
\draw [-, dotted,thick] (A111r) to  (A1111r);
\draw [->] (A1111r) to  (A11111r);

\draw [->] (A13r) to  (A14r);
\draw [->] (A14) to  (A14r);

\draw [->] (A13r) to  (A133r);

\draw [->] (A13) to  (A14);

\end{scope}

\node at (3.25,0) [help](A2) [label=90: $\frm{A}_2$]{};
\node at (3.25,1) [white](rootA2) [ ]{};

\node at (3,1.75) [white](rootA21) [ ]{};
\node at (2.75,2.5) [white](rootA211) [ ]{};

\node at (2.5,3.25) [white](rootA2111) [ ]{};

\begin{scope}[>=stealth, auto]
\draw [->] (rootA2) to  (rootA21);
\draw [-, dotted, thick] (rootA21) to  (rootA211);
\draw [->] (rootA211) to  (rootA2111);
\end{scope}

\node at (4,0) [help](A3) [label=90: $\frm{A}_3$]{};
\node at (4,1) [white](rootA3) [ ]{};
\node at (4,1.75) [white](A31) [ ]{};

\node at (3.7,1.75) [white](rootA31) [ ]{};
\node at (3.45,2.5) [white](rootA311) [ ]{};

\node at (3.2,3.25) [white](rootA3111) [ ]{};

\begin{scope}[>=stealth, auto]
\draw [->] (rootA3) to  (A31);
\draw [->] (rootA3) to  (rootA31);
\draw [-, dotted, thick] (rootA31) to  (rootA311);
\draw [->] (rootA311) to  (rootA3111);
\end{scope}
\node at (4.5,1) [help](ldots) []{$\ldots$};

\node at (5,-0.05) [help](Am1) [label=90: $\frm{A}_{m+1}$]{};
\node at (5,1) [white](rootAm1) [ ]{};
\node at (5,2) [white](Am11) [ ]{};
\node at (5,3) [white](Am12) [ ]{};
\node at (5,3.75) [white](Am13) [ ]{};
\node at (4.7,1.75) [white](rootAm11) [ ]{};
\node at (4.45,2.5) [white](rootAm111) [ ]{};

\node at (4.2,3.25) [white](rootAm1111) [ ]{};

\begin{scope}[>=stealth, auto]
\draw [->] (rootAm1) to  (Am11);
\draw [->] (rootAm1) to  (rootAm11);
\draw [-, dotted, thick] (rootAm11) to  (rootAm111);
\draw [-, dotted, thick] (Am11) to  (Am12);
\draw [->] (Am12) to  (Am13);

\draw [->] (rootAm111) to  (rootAm1111);
\end{scope}

\node at ( 5.51,4) [help](helpUpDotted) []{};
\node at ( 5.51,0) [help](helpDownDotted) []{};

\begin{scope}[>=stealth, auto]



\draw [ dotted] (helpUpDotted) to  (helpDownDotted);


\end{scope}

\node at (6.55,0) [help](B) [label=90: $\frm{B}$]{};
\node at (6.55,1) [white](rootB) [ ]{};
\node at (6.55,2) [white](B1) [ ]{};
\node at (6.55,3) [white](B2) [ ]{};
\node at (6.55,3.75) [white](B3) [ ]{};
\node at (5.75,3.75) [white](B4) [ ]{};
\node at (6.25,1.75) [white](rootB1) [ ]{};
\node at (6,2.5) [white](rootB11) [ ]{};

\node at (5.75,3.25) [white](rootB111) [ ]{};

\begin{scope}[>=stealth, auto]
\draw [->] (rootB) to  (B1);
\draw [->] (rootB) to  (rootB1);
\draw [-, dotted, thick] (rootB1) to  (rootB11);
\draw [-, dotted, thick] (B1) to  (B2);
\draw [->] (B2) to  (B3);
\draw [->] (B3) to  (B4);
\draw [->] (rootB11) to  (rootB111);
\end{scope}



\node at (8.8,-0.1) [help](mA1) [label=90: $\mods{\frm A}_1$]{};
\node at (8.8,1) [white](mrootA1) [label=180:$\triangleright$ ]{};

\node at (8.3,1.5) [white](mrootA11) [ ]{};
\node at (7.8,2) [white](mrootA111) [ ]{};

\node at (7.3,2.5) [black](mrootA1111) [ ]{};

\node at (8.8,2) [white](mA11) [ ]{};

\node at (8.8,3) [white](mA13) [ ]{};

\node at (8.8,3.75) [white](mA14) [ ]{};

\node at (8.4,2) [white](mA11r) [ ]{};
\node at (8,2.5) [white](mA111r) [ ]{};
\node at (7.55,3) [white](mA1111r) [ ]{};
\node at (7.05,3.5) [black](mA11111r) [ ]{};

\node at (8.1,3) [white](mA13r) [ ]{};
\node at (7.4,3.7) [black](mA133r) [ ]{};
\node at (7.8,3.75) [rect](mA14r) [ ]{};

\begin{scope}[>=stealth, auto]
\draw [->] (mrootA1) to  (mA11);
\draw [->] (mrootA1) to  (mA11r);
\draw [->] (mrootA1) to  (mrootA11);
\draw [-, dotted, thick] (mrootA11) to  (mrootA111);
\draw [->] (mrootA111) to  (mrootA1111);

\draw [->] (mA11r) to  (mA111r);
\draw [-, dotted, thick] (mA11r) to  (mA13r);
\draw [-, dotted, thick] (mA11) to  (mA13);
\draw [-, dotted,thick] (mA111r) to  (mA1111r);
\draw [->] (mA1111r) to  (mA11111r);

\draw [->] (mA13r) to  (mA14r);
\draw [->] (mA14) to  (mA14r);

\draw [->] (mA13r) to  (mA133r);

\draw [->] (mA13) to  (mA14);

\end{scope}

\node at (10.05,-0.1) [help](mA2) [label=90: $\mods{\frm A}_2$]{};
\node at (10.05,1) [white](mrootA2) [label=180:$\triangleright$ ]{};

\node at (9.8,1.75) [white](mrootA21) [ ]{};
\node at (9.55,2.5) [white](mrootA211) [ ]{};

\node at (9.3,3.25) [black](mrootA2111) [ ]{};

\begin{scope}[>=stealth, auto]
\draw [->] (mrootA2) to  (mrootA21);
\draw [-, dotted, thick] (mrootA21) to  (mrootA211);
\draw [->] (mrootA211) to  (mrootA2111);
\end{scope}

\node at (11,-0.1) [help](mA3) [label=90:$\mods{\frm A}_3$ ]{};
\node at (11,1) [white](mrootA3) [label=180:$\triangleright$ ]{};
\node at (11,1.75) [white](mA31) [ ]{};

\node at (10.7,1.75) [white](mrootA31) [ ]{};
\node at (10.45,2.5) [white](mrootA311) [ ]{};

\node at (10.2,3.25) [black](mrootA3111) [ ]{};

\begin{scope}[>=stealth, auto]
\draw [->] (mrootA3) to  (mA31);
\draw [->] (mrootA3) to  (mrootA31);
\draw [-, dotted, thick] (mrootA31) to  (mrootA311);
\draw [->] (mrootA311) to  (mrootA3111);
\end{scope}
\node at (11.5,1) [help](mldots) []{$\ldots$};

\node at (12,-0.15) [help](mAm1) [label=90: $\mods{\frm A}_{m+1}$]{};
\node at (12,1) [white](mrootAm1) [label=0:$\triangleleft$ ]{};
\node at (12,2) [white](mAm11) [ ]{};
\node at (12,3) [white](mAm12) [ ]{};
\node at (12,3.75) [white](mAm13) [ ]{};
\node at (11.7,1.75) [white](mrootAm11) [ ]{};
\node at (11.45,2.5) [white](mrootAm111) [ ]{};

\node at (11.2,3.25) [black](mrootAm1111) [ ]{};

\begin{scope}[>=stealth, auto]
\draw [->] (mrootAm1) to  (mAm11);
\draw [->] (mrootAm1) to  (mrootAm11);
\draw [-, dotted, thick] (mrootAm11) to  (mrootAm111);
\draw [-, dotted, thick] (mAm11) to  (mAm12);
\draw [->] (mAm12) to  (mAm13);

\draw [->] (mrootAm111) to  (mrootAm1111);
\end{scope}

\node at ( 12.6,4) [help](mhelpUpDotted) []{};
\node at ( 12.6,0) [help](mhelpDownDotted) []{};

\begin{scope}[>=stealth, auto]



\draw [ dotted] (mhelpUpDotted) to  (mhelpDownDotted);


\end{scope}

\node at (13.7,-0.05) [help](mB) [label=90: $\mods{\frm B}$]{};
\node at (13.7,1) [white](mrootB) [ label=180:$\triangleright$]{};
\node at (13.7,2) [white](mB1) [ ]{};
\node at (13.7,3) [white](mB2) [ ]{};
\node at (13.7,3.75) [white](mB3) [ ]{};
\node at (12.9,3.75) [rect](mB4) [ ]{};
\node at (13.4,1.75) [white](mrootB1) [ ]{};
\node at (13.15,2.5) [white](mrootB11) [ ]{};

\node at (12.9,3.25) [black](mrootB111) [ ]{};

\begin{scope}[>=stealth, auto]
\draw [->] (mrootB) to  (mB1);
\draw [->] (mrootB) to  (mrootB1);
\draw [-, dotted, thick] (mrootB1) to  (mrootB11);
\draw [-, dotted, thick] (mB1) to  (mB2);
\draw [->] (mB2) to  (mB3);
\draw [->] (mB3) to  (mB4);
\draw [->] (mrootB11) to  (mrootB111);
\end{scope}

\node at ( 6.8,-0.2) [help](mhelpDown) []{};
\node at ( 6.8,4.25) [help](mhelpUp) []{};
\node at ( 0,4.25) [help](mhelpleftUp) []{};
\node at ( 0,-0.2) [help](mhelpleftDown) []{};
\node at ( 14,4.25) [help](mhelpRightUp) []{};
\node at ( 14,-0.2) [help](mhelpRightDown) []{};

\begin{scope}[>=stealth, auto]

\draw  (mhelpUp) to  (mhelpDown);
\draw  (mhelpleftUp) to  (mhelpleftDown);

\draw  (mhelpRightUp) to  (mhelpRightDown);

\draw  (mhelpUp) to  (mhelpRightUp);
\draw  (mhelpUp) to  (mhelpleftUp);
\draw   (mhelpDown) to (mhelpleftDown);
\draw   (mhelpDown) to (mhelpRightDown);
\end{scope}

\end{tikzpicture}
\end{center}
\caption{The frames ${\frm A}_{1}$, $\ldots$, ${\frm A}_{m+1}$ and ${\frm B}$ and the pointed models based on them.}
\label{fig:nsmallerthanm}
\end{figure}

\noindent{\sc selection of the models on the right:} In this case, Hercules must choose his models according to the following.

\begin{lemma}\label{lemA1BTrans}
In any winning strategy for Hercules for an $( \lang_{\pd},\langle \cls{L},\cls{R}\rangle)$-{\fsgf} in which $\frm{A}_1\in \cls{L}$ and $\frm{B}\in \cls{R}$, Hercules picks a pointed model $(\mods{\frm B},\point)$ based on the lowest point in $\frm{B}$, and assigns different valuations to the two dead-end points of $\frm B$.
\end{lemma}

\proof
Hercules is not going to select a pointed model that is not based on the lowest point in $\frm{B}$
because the Hydra can always reply with a bisimilar pointed model based on $\frm{A}_1$. Similarly, if Hercules assigns the same valuation to the two dead-ends the Hydra can choose a bisimilar model based on $\frm A_1$ by copying the valuations from the hypothenuse onto all paths of length $n$, and copying the valuations from the legs onto the path of length $m$; since the valuations coincide on the end-points, there is no clash at the top left of the triangle.
\endproof

To indicate that the two end-points of ${\frm B}$ receive different valuations, we have drawn one of them black 
while the other is shaped as a rectangle. The literals true in the rest of the points are immaterial. Thus, Hercules
constructs the pointed model $(\mods{\frm{B}}, \triangleright)$ shown in the right rectangle in Figure~\ref{fig:nsmallerthanm}.
\medskip

\noindent{\sc selection of models on the left:}
The Hydra replies with the pointed models shown on the left of the dotted line in the right rectangle in Figure~\ref{fig:nsmallerthanm}. 
 The pointed model based on $\frm{A}_1$ is defined so that the set of literals true in the points on a diagonal path that shares points with the hypotenuse but do not coincide with it copy the respective sets of literals true in the points of  the diagonal path in $\frm{B}$.

The models $\frm{A}_i$ for $2\leq i\leq m+1$ receive valuations so that their diagonal paths coincide  with the diagonal path in the model $\frm{B}$
whereas their vertical paths are  `initial segments' of the vertical path in
$\frm{B}$, i.e., the lowest  point in any $\frm{A}_i$  for $2\leq i\leq m+1$ and the lowest  point in $\frm{B}$ satisfy the same literals and similarly for their vertical successors.
As before, if the Hydra chooses her models in this way, we say that she {\em mimics} Hercules' choice.
\medskip

\noindent{\sc formula size game on models:}
We consider the {\fsgm} starting with  
$(\mods{\frm A}_1, \triangleright),\ldots, ( \mods {\frm A}_{m+1}, \triangleright)$ on the left and $(\mods{\frm B}, \triangleright)$
on the right. These lemmas are analogous to those in Section \ref{secDens}.

\begin{lemma}\label{lemTransHyp}
Let $\cls L$, $\cls R$ be classes of models so that Hercules has a winning strategy for the $( \lang_{\pd},\langle \cls{L},\cls{R}\rangle)$-{\fsgm}.
Let $T$ be any closed game in which the Hydra played greedily and $\eta$ be a node on which Hercules played a $\pd$-move.
\begin{enumerate}
\item 
If $(\mods{\frm{A}}_1, \triangleright)\in \lft (\eta)$ and $(\mods{\frm{B}}, \triangleright)\in \rgt(\eta)$, then
he picked a pointed model based on a point that lies on the hypotenuse of $\mods{\frm{A}}_1$.

\item  If for some $i\in [3,m+1]$ we have that $(\mods{\frm{A}}_i, \triangleright) \in \lft(\eta)$ and $(\mods{\frm{B}}, \triangleright)\in \rgt(\eta)$, then
he picked the rightmost daughter as a successor of $(\mods{\frm A }_i, \triangleright)$.

\end{enumerate}
\end{lemma}

\proof
Both items hold because if Hercules picked a different point, the Hydra replied with the same point in $\mods{\frm{B}}$.
In either case we obtain bisimilar models on each side, which by Corollary~\ref{lemHercLose} means that Hercules cannot win.
\endproof

\begin{lemma}\label{lemTransNoModA}
Suppose that $\cls L$ and $\cls R$ are classes of models and Hercules has a winning strategy for the $( \lang_{\pd},\langle \cls{L},\cls{R}\rangle)$-{\fsgm}.
Suppose that $T$ is a closed game tree, the Hydra played greedily, and $\eta$ is a node of $T$.
\begin{enumerate}

\item If $(\mods{\frm A}_1, \triangleright)\in \lft(\eta)$  and $(\mods{\frm B}, \triangleright)\in \rgt(\eta)$, then Hercules did not play a $\Box$-move at $\eta$.

\item If $(\mods{\frm A}_2,\triangleright) \in \lft(\eta)$ and $(\mods{\frm B}, \triangleright) \in \rgt(\eta)$,
 then Hercules did not play a $\pd$-move at $\eta$.

\end{enumerate}
\end{lemma}

\proof
The first item is immedate from the fact that if Hercules played a $\Box$-move, the Hydra can reply with 
the same point in $\mods{\frm A}_1$, and similarly in the second case the Hydra would reply with the same pointed model based on $\mods{\frm B}$.
\endproof

As was the case for the generalized density axioms, Hercules must play at least one $\vee$-move to separate $\mods{\frm A}_1$ from $\mods{\frm A}_2$.

\begin{proposition}\label{propTransBoundsA}
Let $\cls L$ and $\cls R$ be classes of models such that Hercules has a winning strategy for the $( \lang_{\pd},\langle \cls{L},\cls{R}\rangle)$-{\fsgm}.
Let $T$ be a closed game tree in which the Hydra played greedily.
\begin{enumerate}

\item\label{itTransDisA} If $ (\frm{A}_1, \triangleright), (\frm{A}_2,\triangleright) \in \cls{L}$ and $(\frm{B}, \triangleright)\in \cls{R}$, then Hercules made
at least one $\vee$-move during the game.

\item \label{itTransDiamA} If $(\mods{\frm{A}}_1, \triangleright)\in \cls{L}$, then $T$ has at least $n$ nested $\pd$-moves and at least one literal move.

\item\label{itTransBoxA}
If $\{(\mods{\frm A}_2, \triangleright),\ldots, (\mods{\frm A}_{m+1},\triangleright)\}\subseteq \cls{L}$, then $T$ has at least $m$ $\Box$-moves.

\end{enumerate}

\end{proposition}

\proof
The proof of the first item is analogous to that of Proposition \ref{propDensBounds}.\ref{itDensDisj}, except that it uses Lemma \ref{lemTransNoModA}, and the proof of the third item is essentially the same as the proof of Proposition \ref{propDensBounds}.\ref{itDensBox}. Thus we focus on the second item.

Since $(\mods{\frm A}_1,\triangleright)$ and $(\mods{\frm B},\triangleright)$ satisfy the 
same literals and since $\vee$- and $\wedge$-moves lead to at least one new game-position  in which   $(\mods{\frm A}_1,\triangleright)$
is on the left and $(\mods{\frm B},\triangleright)$ is on the right, Hercules must perform a $\pd$-move in a position in which  $(\mods{\frm A}_1,\triangleright)$
is on the left and $(\mods{\frm B},\triangleright)$ is on the right. It follows from Lemma \ref{lemTransHyp},  that he is going to pick the immediate successor along the hypotenuse
of  $\mods{\frm A}_1$. The Hydra replies, with among others, the immediate successor along the diagonal path  in $\mods{\frm B}$.
Since the new pointed models satisfy the same literals, Hercules cannot play a literal-move; moreover,
$\vee$- and $\wedge$-moves  lead to at least one new game position that is essentially the same as the previous one. If he decided to play a $\Box$-move and
picked a pointed model based on a point along the diagonal path in $\mods{\frm B}$, the Hydra will reply with the same point along a path 
that is different from the hypotenuse because such paths are always available. Hence, he must make at least $n-1$ subsequent $\pd$-moves
to reach the point in which the hypotenuse of  $\mods{\frm A}_1$ and its horizontal leg meet.
Finally, at this point Hercules must play a literal, as this is the only move that will lead to a closed game-tree.
\endproof

\noindent With this we conclude the proof of Theorem \ref{theoTransfer} in the case $0<n<m$.

\proof
Similar to the proof for the case $0<m<n$, except that we use the classes $\cls A$, $\cls B$ of Figure \ref{fig:nsmallerthanm} and Proposition \ref{propTransBoundsA}.
\endproof
%
%
%
%
%
%

Now we proceed to proving Theorem \ref{theoTransfer} in the cases where one of the parameters is zero.
\subsection{The generalized reflexivity axioms}

\begin{figure}
\begin{center}
\begin{tikzpicture}
[
help/.style={circle,draw=white!100,fill=white!100,thick, inner sep=0pt,minimum size=1mm},
white/.style={circle,draw=black!100,fill=white!100,thick, inner sep=0pt,minimum size=2mm},
black/.style={circle,draw=black!100,fill=black!100,thick, inner sep=0pt,minimum size=2mm},
xscale=0.9,
yscale=0.9
]

\node at (1,0.05) [help](A) [label=90: $\frm{A}_1$]{};
\node at (1,1) [white](loopA) [ ]{};

\begin{scope}[>=stealth, auto]
\draw [->] (loopA) [out=50, in=130, loop]to  (loopA);
\end{scope}

\node at (3.5,0.05) [help](B) [label=90: $\frm{A}_2$]{};
\node at (3.5,1) [white](rootB) []{};
\node at (3.5,2) [white](1B) [ ]{};

\node at (3.5,3) [white](2B) [ ]{};
\node at (3.5,4) [white](3B) [ ]{};

\node at (2.6,1)[help](Blengthdown1)[]{};
\node at (2.8,1)[help](Blengthdown2)[]{};
\node at (3,1)[help](Blengthdown3)[]{};

\node at (2.6,4)[help](Blengthup1)[]{};
\node at (2.8,4)[help](Blengthup2)[]{};
\node at (3,4)[help](Blengthup3)[]{};

\node at (4.5,2.5) [white](loopB) [ ]{};


\begin{scope}[>=stealth, auto]
\draw [->] (rootB) to  (1B);

\draw [-,dotted] (1B) to  (2B);
\draw [->] (2B) to  (3B);
\draw [->, bend right=30] (3B) to (rootB);

\draw[-](Blengthdown1) to (Blengthdown3);
\draw[-](Blengthup1) to (Blengthup3);
\draw [<->, dashed] (Blengthdown2) to node{$n-1$} (Blengthup2);

\draw [->] (rootB) to  (loopB);

\draw [->] (1B) to  (loopB);
\draw [->] (2B) to  (loopB);
\draw [->] (3B) to  (loopB);
\draw [->] (loopB) [out=320, in=40, loop]to  (loopB);
\end{scope}

\node at ( 5.5,4) [help](helpUpDotted) []{};
\node at ( 5.5,0) [help](helpDownDotted) []{};

\begin{scope}[>=stealth, auto]



\draw [ dotted] (helpUpDotted) to  (helpDownDotted);


\end{scope}

\node at (6,0.1) [help](C) [label=90: $\frm{B}$]{};
\node at (6,1) [white](rootC) []{};
\node at (6,2) [white](loopC) [ ]{};


\begin{scope}[>=stealth, auto]
\draw [->] (rootC) to  (loopC);
\draw [->] (loopC) [out=50, in=130, loop]to  (loopC);

\end{scope}
\node at ( 6.75,-0.25) [help](mhelpDown) []{};
\node at ( 6.75,4.5) [help](mhelpUp) []{};
\node at ( 0.5,4.5) [help](mhelpleftUp) []{};
\node at ( 0.5,-0.25) [help](mhelpleftDown) []{};
\node at ( 13.5,4.5) [help](mhelpRightUp) []{};
\node at ( 13.5,-0.25) [help](mhelpRightDown) []{};

\begin{scope}[>=stealth, auto]

\draw  (mhelpUp) to  (mhelpDown);
\draw  (mhelpleftUp) to  (mhelpleftDown);

\draw  (mhelpRightUp) to  (mhelpRightDown);

\draw  (mhelpUp) to  (mhelpRightUp);
\draw  (mhelpUp) to  (mhelpleftUp);
\draw   (mhelpDown) to (mhelpleftDown);
\draw   (mhelpDown) to (mhelpRightDown);
\end{scope}

\node at (7.5,0) [help](mA) [label=90: $\mods{\frm A}_1$]{};
\node at (7.5,1) [white](mloopA) [label=0: $\triangleleft$ ]{};

\begin{scope}[>=stealth, auto]
\draw [->] (mloopA) [out=50, in=130, loop]to  (mloopA);
\end{scope}

\node at (10,0) [help](mB) [label=90:$\mods{\frm A}_2$]{};
\node at (10,1) [black](mrootB) [label=0: $\triangleleft$]{};
\node at (10,2) [white](m1B) [ ]{};

\node at (10,3) [white](m2B) [ ]{};
\node at (10,4) [white](m3B) [ ]{};

\node at (9.1,1)[help](mBlengthdown1)[]{};
\node at (9.3,1)[help](mBlengthdown2)[]{};
\node at (9.5,1)[help](mBlengthdown3)[]{};

\node at (9.1,4)[help](mBlengthup1)[]{};
\node at (9.3,4)[help](mBlengthup2)[]{};
\node at (9.5,4)[help](mBlengthup3)[]{};

\node at (11,2.5) [white](mloopB) [ ]{};


\begin{scope}[>=stealth, auto]
\draw [->] (mrootB) to  (m1B);

\draw [-,dotted] (m1B) to  (m2B);
\draw [->] (m2B) to  (m3B);
\draw [->, bend right=30] (m3B) to (mrootB);

\draw[-](mBlengthdown1) to (mBlengthdown3);
\draw[-](mBlengthup1) to (mBlengthup3);
\draw [<->, dashed] (mBlengthdown2) to node{$n-1$} (mBlengthup2);

\draw [->] (mrootB) to  (mloopB);

\draw [->] (m1B) to  (mloopB);
\draw [->] (m2B) to  (mloopB);
\draw [->] (m3B) to  (mloopB);
\draw [->] (mloopB) [out=320, in=40, loop]to  (mloopB);
\end{scope}

\node at ( 12,4) [help](mhelpUpDotted) []{};
\node at ( 12,0) [help](mhelpDownDotted) []{};

\begin{scope}[>=stealth, auto]



\draw [ dotted] (mhelpUpDotted) to  (mhelpDownDotted);


\end{scope}

\node at (12.5,0.05) [help](mC) [label=90: $\mods{\frm B}$]{};
\node at (12.5,1) [black](mrootC) [label=0: $\triangleleft$]{};
\node at (12.5,2) [white](mloopC) [ ]{};


\begin{scope}[>=stealth, auto]
\draw [->] (mrootC) to  (mloopC);
\draw [->] (mloopC) [out=50, in=130, loop]to  (mloopC);

\end{scope}

\end{tikzpicture}
\end{center}
\caption{The frames  $\frm{A}_1$, $\frm{A}_2$ and $\frm{B}$ and the pointed models based on them. }
\label{fig:m0}
\end{figure}
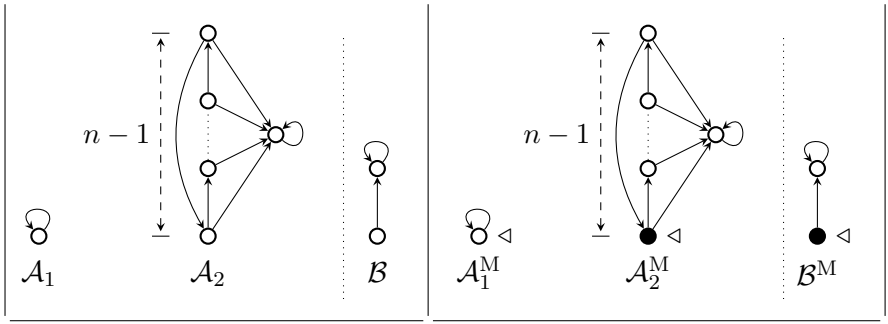

Recall that we write {\em $n$-reflexivity} instead of {$(0,n)$-transfer.}
In order to prove that Theorem \ref{theoTransfer} holds in this case, we consider a $( \lang _{\pd},\langle \cls{A},\cls{B}\rangle)$-{\fsgf} where 
$\cls{A}=\{\frm{A}_1, \frm{A}_2\}$ and $\cls{B}$ contains a single element $\frm{B}$.
These frames are shown in the left rectangle in Figure~\ref{fig:m0} and separated by the dotted line.
The ``highest'' point in $\frm{A}_2$ can be reached in $n-1$ relation steps from the lowest one and then we can return back 
to the latter in one additional relation step, i.e, the points in $\frm{A}_2$ that are different from the reflexive one form a cycle of length $n$.
It is immediate that $p\rightarrow\pd^np$ is valid on both $\frm{A}_1$ and $\frm{A}_2$ and not valid on $\frm{B}$.

Next we  study Hercules' possible strategies. We begin with his choice of models on the right. 
\medskip

\noindent{\sc selection of the pointed models on the right:}
If Hercules is to win the formula-complexity game, he must choose his models  in a specific way.

\begin{lemma}\label{lemA1B}
In any winning strategy for Hercules for an $( \lang_{\pd},\langle \cls{L},\cls{R}\rangle)$-{\fsgf} in which $\frm{A}_1\in \cls{L}$ and $\frm{B}\in \cls{R}$,
\begin{enumerate}

\item\label{itA1BOne} Hercules chooses the valuation on $\frm B$ so that at least one literal is true in one point but not on the other, and

\item\label{itA1BTwo} he picks the pointed model based on the irreflexive point in $\frm{B}$. 

\end{enumerate}

\end{lemma}

\putaway{
\proofLet $a$ be the single point in $\frm A_1$, $i$ be the irreflexive point in $\frm B$ and $r$ be the reflexive point, and $V_\frm B$ be the valuation on $\frm B$ that Hercules chooses. Then Hydra can reply by setting $V_{\frm A_1} (a) = V_\frm B(r)$. Then if $V_\frm B (i) = V_\frm B (r)$ or if Hercules chooses $r$, then there would
be two bisimilar pointed models, one on the left and the other on the right, in the subsequent formula-complexity game, so that by Corollary~\ref{lemHercLose} Hydra has a winning strategy on the subsequent $(\Mods{\cls A},\Mods{\cls B})$-{\fsgm}. \endproof
}
 
 The pointed model based on $\frm{B}$ and its irreflexive point chosen by Hercules is shown in the right half of Figure~\ref{fig:m0}. 
We indicate that the two points in $\frm{B}$ satisfy different sets of literals by making one of them black and the other white.
\medskip

\noindent{\sc selection of the pointed models on the left:}
The Hydra can reply with the pointed models shown on the left of the dotted line in the right half in Figure~\ref{fig:m0}. 
She selects these pointed models so that two points in any two models satisfy the same set of literals iff they have  the same colour. As usual, we say that she {\em mimics} Hercules if she chooses her pointed models in this way.
\medskip

\noindent{\sc formula size game on models:}
Let us consider now the {\fsgm} starting with  
$(\mods{\frm A}_1, \triangleright), (\mods{\frm A}_2, \triangleright)$ on the left and $(\mods{\frm B}, \triangleright)$
on the right.
We first note some restrictions on Hercules's modal moves. 
The following can be seen by observing that playing otherwise would produce bisimilar pointed models on each side.

\begin{lemma}\label{lem:A1Bbox} Let $\cls L$, $\cls R$ be classes of models so that Hercules has a winning strategy for the  $( \lang_{\pd},\langle \cls{L},\cls{R}\rangle)$-{\fsgm} and $T$ a closed game tree in which the Hydra played greedily.

\begin{enumerate}

\item If there is a game position $\eta$ in which any pointed model based on either $\mods{\frm A}_1$
or $\mods{\frm{A}}_2$ is on the left and any pointed model based on $\mods{\frm{B}}$ is on the right, then Hercules did not play a $\Box$-move at $\eta$.

\item If there is a game position $\eta$ in which  $(\mods{\frm A}_1, \triangleright)$ is on the left and a pointed model
based on $\mods{\frm{B}}$ is on the right, then Hercules did not play a $\pd$-move at $\eta$.

\end{enumerate}

\end{lemma}

\putaway{
\proof
For the first item, we observe that, when making such a move, Hercules must chose the 
reflexive point in $ \mods{\frm{B}}$ to which the Hydra will reply with the bisimilar reflexive point in $\mods{\frm{A}}_1$ 
or the bisimilar reflexive point in $\mods{\frm{A}}_2$, respectively, which by Corollary~\ref{lemHercLose} implies that Hercules cannot win.

As far as the second item is concerned, note that
if Hercules makes a $\pd$-move, he must  remain
in the reflexive point in $\mods{\frm{A}}_1$ but the Hydra is going to reply with the bisimilar reflexive point in $\mods{\frm{B}}$.
\endproof
}

From this it is easy to see that Hercules must play at least one variable.

\begin{lemma}\label{lem:A1Bvar} Suppose that $\cls L$, $\cls R$ are classes of models and that Hercules has a winning strategy for the  $( \lang_{\pd},\langle \cls{L},\cls{R}\rangle)$-{\fsgm}. Let $T$ be a closed game tree in which the Hydra played greedily and such that there is a position $\eta$ in which $(\mods{\frm A}_1,\point)$ is on the left and $(\mods{\frm B},\point)$ is on the right. Then, the number of literal moves in $T$ is at least one.
\end{lemma}

\proof
By Lemma \ref{lem:A1Bbox} Hercules cannot play any $\pd$- or $\Box$-moves, and $\wedge$- or $\vee$-moves result in at least one new position with both of these pointed models. Since Hercules cannot play $\bot$ or $\top$, he must use at least one variable.
\endproof

\noindent With this we are ready to prove Theorem \ref{theoTransfer} in the case where $m=0$.

\proof
Let $\cls{A}$ and $\cls{B}$ be as depicted in the left rectangle in Figure~\ref{fig:m0}; since the frames of $\cls A$ are $n$-reflexive but the ones in $\cls B$ are not, by Theorem \ref{thrm: validityGames} it suffices to show that the Hydra can play so that any closed game tree has at least one $\vee$-move, one literal move, and modal depth at least $n$.

Let $\Mods{\cls B} = \{ ( \mods{\frm B},\point_\frm B) \}$ be the singleton set of pointed models chosen by Hercules, which by Lemma \ref{lemA1B} must be so that the top and bottom points have different valuations, and let Hydra choose $\Mods{\cls A}$ as depicted in the right-hand side of Figure \ref{fig:m0}.
Lemma~\ref{lem:A1Bbox} implies  that Hercules cannot begin the {\fsgm} starting with
$(\mods{\frm A}_1, \triangleright), (\mods{\frm A}_2, \triangleright)$ on the left and $(\mods{\frm B}, \triangleright)$ on the right by playing either a $\pd$- or a $\Box$-move. Playing an $\wedge$-move will result in at least one new
position that is the same as the previous one. Therefore, Hercules must play an $\vee$-move
and he and the Hydra will have to compete in  two new sub-games:
the first one starting
with $(\mods{\frm{A}}_1, \triangleright)$ on the left and $(\mods{\frm{B}}, \triangleright)$ on the right
while the second starts with $(\mods{\frm{A}}_2,\triangleright)$ on the left and    $(\mods{\frm{B}}, \triangleright)$ on the right.

By Lemma \ref{lem:A1Bvar} he can win the former only by playing a literal-move whereas the latter can be won only by playing
a sequence of $n$ $\pd$-moves that  must be made in order to  perform a cycle leading back to 
the black point in $\frm{A}_2$, giving us at least $n$ ocurrences of $\pd$ and modal depth at least $n$.
We can then use Theorem \ref{thrm: validityGames} to conclude that $\overline p \vee \pd^n p $ is absolutely minimal.
\endproof

\subsection{The generalized recurrence axioms}

Now we treat the $m$-recurrence axioms, where $n = 0$.
This time Hercules and the Hydra play a $( \lang_{\pd},\langle \cls{A},\cls{B}\rangle)$-{\fsgf} where
$\cls{A}=\{\frm{A}_1, \ldots \frm{A}_{m+1}\}$ while $\cls{B}$ contains a single element $\frm{B}$, as depicted in the left rectangle in Figure~\ref{fig:n0}.
\begin{figure} 
\begin{center}
\begin{tikzpicture}
[
help/.style={circle,draw=white!100,fill=white!100,thick, inner sep=0pt,minimum size=1mm},
white/.style={circle,draw=black!100,fill=white!100,thick, inner sep=0pt,minimum size=2mm},
black/.style={circle,draw=black!100,fill=black!100,thick, inner sep=0pt,minimum size=2mm},
xscale=0.95,
yscale=0.95
]


\node at (0.5,0.05) [help](A1) [label=90: $\frm{A}_1$]{};
\node at ( 0.5,1) [white](loopA1) [ ]{};

\begin{scope}[>=stealth, auto]

\draw [->] (loopA1) [out=50, in=130, loop]to  (loopA1);

\end{scope}

\node at (1.25,0.05) [help](A2) [label=90: $\frm{A}_2$]{};
\node at ( 1.25,1) [white](A2) [ ]{};

\node at (2,0.05) [help](A3) [label=90: $\frm{A}_3$]{};
\node at ( 2,1) [white](A31) [ ]{};
\node at ( 2,2) [white](A32) [ ]{};
\begin{scope}[>=stealth, auto]

\draw [->] (A31) to  (A32);

\end{scope}

\node at ( 2.5,1) [help](ldots) [ ]{$\ldots$};

\node at (3,0.05) [help](Amplus1) [label=90: $\frm{A}_{m+1}$]{};
\node at ( 3,1) [white](A41) [ ]{};
\node at ( 3,2) [white](A42) [ ]{};
\node at ( 3,2.6) [help](A4vdots) [ ]{$\vdots$};
\node at ( 3,3) [white](A43) [ ]{};
\node at ( 3,4) [white](A44) [ ]{};
\begin{scope}[>=stealth, auto]

\draw [->] (A41) to  (A42);
\draw [->] (A43) to  (A44);
\end{scope}

\node at (3.2,1)[help](A4lengthdown1)[]{};
\node at (3.4,1)[help](A4lengthdown2)[]{};
\node at (3.6,1)[help](A4lengthdown3)[]{};

\node at (3.2,4)[help](A4lengthup1)[]{};
\node at (3.4,4)[help](A4lengthup2)[]{};
\node at (3.6,4)[help](A4lengthup3)[]{};
\begin{scope}[>=stealth, auto]
\draw[-](A4lengthdown1) to (A4lengthdown3);
\draw[-](A4lengthup1) to (A4lengthup3);
\draw [<->, dashed, swap] (A4lengthdown2) to node{$m-1$} (A4lengthup2);
\end{scope}

\node at ( 4.6,4) [help](helpUpDotted) []{};
\node at ( 4.6,0) [help](helpDownDotted) []{};
\begin{scope}[>=stealth, auto]



\draw [ dotted] (helpUpDotted) to  (helpDownDotted);


\end{scope}


\node at (5,0.08) [help](B) [label=90: $\frm{B}$]{};
\node at ( 5,1) [white](rootB) [ ]{};
\node at ( 5,2) [white](loopB) [ ]{};

\begin{scope}[>=stealth, auto]

\draw [->] (rootB) to  (loopB);
\draw [->] (loopB) [out=50, in=130, loop]to  (loopB);

\end{scope}

\node at ( 5.5,-0.06) [help](mhelpDown) []{};
\node at ( 5.5,4.5) [help](mhelpUp) []{};
\node at ( 0,4.5) [help](mhelpleftUp) []{};
\node at ( 0,-0.06) [help](mhelpleftDown) []{};
\node at ( 12,4.5) [help](mhelpRightUp) []{};
\node at ( 12,-0.06) [help](mhelpRightDown) []{};

\begin{scope}[>=stealth, auto]

\draw  (mhelpUp) to  (mhelpDown);
\draw  (mhelpleftUp) to  (mhelpleftDown);

\draw  (mhelpRightUp) to  (mhelpRightDown);

\draw  (mhelpUp) to  (mhelpRightUp);
\draw  (mhelpUp) to  (mhelpleftUp);
\draw   (mhelpDown) to (mhelpleftDown);
\draw   (mhelpDown) to (mhelpRightDown);
\end{scope}


\node at (6,0) [help](mA1) [label=90:  $\mods{\frm A}_1$]{};
\node at ( 6,1) [white](mloopA1) [label=180:$\triangleright$ ]{};

\begin{scope}[>=stealth, auto]

\draw [->] (mloopA1) [out=50, in=130, loop]to  (mloopA1);

\end{scope}

\node at (6.75,0) [help](mA2) [label=90:  $\mods{\frm A}_2$]{};
\node at ( 6.75,1) [black](mA2) [ label=180:$\triangleright$]{};

\node at (7.5,0) [help](mA3) [label=90:  $\mods{\frm A}_3$]{};
\node at ( 7.5,1) [black](mA31) [label=180:$\triangleright$ ]{};
\node at ( 7.5,2) [white](mA32) [ ]{};
\begin{scope}[>=stealth, auto]

\draw [->] (mA31) to  (mA32);

\end{scope}

\node at ( 8,1) [help](mldots) [ ]{$\ldots$};

\node at (9,0) [help](mAmplus1) [label=90:  $\mods{\frm A}_{m+1}$]{};
\node at ( 9,1) [black](mA41) [label=180:$\triangleright$ ]{};
\node at ( 9,2) [white](mA42) [ ]{};
\node at ( 9,2.6) [help](mA4vdots) [ ]{$\vdots$};
\node at ( 9,3) [white](mA43) [ ]{};
\node at ( 9,4) [white](mA44) [ ]{};
\begin{scope}[>=stealth, auto]

\draw [->] (mA41) to  (mA42);
\draw [->] (mA43) to  (mA44);
\end{scope}

\node at (9.2,1)[help](mA4lengthdown1)[]{};
\node at (9.4,1)[help](mA4lengthdown2)[]{};
\node at (9.6,1)[help](mA4lengthdown3)[]{};

\node at (9.2,4)[help](mA4lengthup1)[]{};
\node at (9.4,4)[help](mA4lengthup2)[]{};
\node at (9.6,4)[help](mA4lengthup3)[]{};
\begin{scope}[>=stealth, auto]
\draw[-](mA4lengthdown1) to (mA4lengthdown3);
\draw[-](mA4lengthup1) to (mA4lengthup3);
\draw [<->, dashed, swap] (mA4lengthdown2) to node{$m-1$} (mA4lengthup2);
\end{scope}

\node at ( 10.75,4) [help](mhelpUpDotted) []{};
\node at ( 10.75,0) [help](mhelpDownDotted) []{};
\begin{scope}[>=stealth, auto]



\draw [ dotted] (mhelpUpDotted) to  (mhelpDownDotted);


\end{scope}


\node at (11.6,0.07) [help](mB) [label=90:  $\mods{\frm B}$]{};
\node at ( 11.6,1) [black](mrootB) [label=180:$\triangleright$ ]{};
\node at ( 11.6,2) [white](mloopB) [ ]{};

\begin{scope}[>=stealth, auto]

\draw [->] (mrootB) to  (mloopB);
\draw [->] (mloopB) [out=50, in=130, loop]to  (mloopB);

\end{scope}

\end{tikzpicture}
\end{center}
\caption{The frames  $\frm{A}_1$, $\ldots$, $\frm{A}_{m+1}$ and $\frm B$ and the pointed models based on them. }
\label{fig:n0}
\end{figure}
  For $2\leq i \leq m+1$, each $\frm{A}_i$ is a path 
of $i-2$ relation steps.
Clearly, $\pd^m p\rightarrow p$ is valid in all the frames in $\cls{A}$ and it is not valid in the frame $\frm{B}$.\\

\noindent{\sc selection of the models on the right:}
It follows from Lemma~\ref{lemA1B} that Hercules must pick the pointed model $(\mods{\frm{B}}, \triangleright)$
 shown in the right half of  Figure~\ref{fig:n0}. Again, to indicate that the two points of $\mods{\frm{B}}$ satisfy different sets of literals, we   
colour  one of them black and the other white.
\medskip

\noindent{\sc selection of the pointed models on the left:}
The Hydra replies with the pointed  models shown on the left of the dotted line in the right half of Figure~\ref{fig:n0}. 
Again, she picks these pointed models so that points that satisfy the same set of literals  have  the same colour.
\medskip

\noindent{\sc formula size game on models:}
Let us consider the {\fsgm} starting with the models
$\Mods{\cls A} = \{ (\mods{\frm A}_1, \triangleright),\ldots, (\mods{\frm A}_{m+1}, \triangleright)\}$ on the left and $\Mods{\cls B} = \{(\mods{\frm B}, \triangleright)\}$
on the right.

\begin{lemma}\label{lemRecVee}
 In any closed game tree $T$ for the $(\lang_\pd, \langle \Mods{\cls A},\Mods{\cls B} \rangle )$-{\fsgm} in which the Hydra played greedily, Hercules played at least one $\vee$-move.
\end{lemma}

\proof
Using Lemma~\ref{lem:A1Bbox}, we see that in order to win
a {\fsgm} with a starting position $\eta$ in which $(\mods{\frm{A}}_1, \triangleright)$ is on the left and $(\mods{\frm B}, \triangleright)$ is on the right,
Hercules must not play either a $\pd$- or a $\Box$-move at $\eta$.
On the other hand, for every game in which  there is some $(\mods{\frm A}_i, \triangleright)$ for $2\leq i\leq m+1$ among the pointed models chosen by the Hydra 
 and $(\mods{\frm B}, \triangleright)$ is among the models chosen by Hercules,  if he wants to win the game, then  there is at least one game position  $\nu$ such that   $(\mods{\frm A}_i, \triangleright)$ is on the left and $(\mods{\frm B}, \triangleright)$ is on the right and Hercules played   at least one $\pd$-
or $\Box$-move at $\nu$.
This implies that in any  {\fsgm} with a starting position in which the pointed models selected by the Hydra
 are on the left and $(\mods{\frm B}, \triangleright)$ is on the right, Hercules must play at least one $\vee$
to separate the set of $(\mods{\frm A}_i, \triangleright)$, for $2\leq i$, 
from $(\mods{\frm A}_1, \triangleright)$.
\endproof

\begin{lemma}\label{lem:boxmovesA}
Let $\cls L$, $\cls R$ be classes of models so that Hercules has a winning strategy for the  $( \lang_{\pd},\langle \cls{L},\cls{R}\rangle)$-{\fsgm}.
Let $T$ be a closed game tree in which the Hydra played greedily. If all $(\mods{\frm A}_i,\triangleright)$ for $2\leq i\leq m+1$ are in $\cls{L}$  and $(\mods{\frm B},\triangleright)\in \cls{R}$,
Hercules must have played at least $m$ $\Box$-moves and the modal depth of $T$ must be at least $m$.
\end{lemma}
We omit the proof, which is similar to that of Proposition \ref{propDensBounds}.\ref{itDensBox}.
With this we are ready to prove Theorem~\ref{theoTransfer} for the case where $n=0$.

\proof
Consider the $(\cls A,\cls B)$-{\fsgf} where $\cls A$, $\cls B$ are as depicted in Figure \ref{fig:n0} on the left: by Lemma \ref{lemA1B}, Hercules must choose different valuations for the points of $\cls B$ and choose the bottom point. Let Hydra reply as depicted on the right-hand side of the figure.

By Lemma \ref{lem:A1Bvar}, Hercules must play at least one variable, by Lemma \ref{lemRecVee} he must play at least one $\vee$-move, by Lemma \ref{lem:boxmovesA} he must play at least $m$ $\Box$-moves and modal depth at least $m$ on the resulting {\fsgm}, and we can apply Theorem \ref{thrm: validityGames}.
\endproof

\section{The $\mathsf{S4}$ axiom}\label{secSFour}

Although certain instances of the transfer axioms are studied in isolation, many familiar modal logics are built from combining several basic axioms, possibly (but not necessarily) transfer axioms of several types.
In this and the next section we consider two important examples.
Here we study $\mathsf{S4}$, the modal logic of (finite) preorders.

\begin{theorem}\label{s4axiom}
 The formula $(\overline{p} \wedge \Box\Box\overline{p})\vee\pd p$ is absolutely minimal among the set of 
$\lang_{\pd }$-formulas defining the class of reflexive and transitive frames.
\end{theorem}

It is tempting to think that  Theorem~\ref{s4axiom}  can be proved by modifying slightly  the frames 
depicted in Figure~\ref{fig:nsmallerthanm} for $m=2$ and $n=1$ so as to take care of reflexivity
 and applying more or less the reasoning from 
Sub-section~\ref{sect:trans}. However, a closer look reveals that we have to make sure that Hercules is forced 
to make at least one $\wedge$-move. This means that we must have at least two models on the right for the model equivalence game.
Additionally, there are no guarantees that Lemma~\ref{lemTransNoModA} will remain true if we make the relations  in the frames of $\cls{A}$ reflexive  (in fact it does not). Nevertheless, we can make this strategy work by taking some extra care. 

\begin{figure} [htbp]
\begin{center}
\begin{tikzpicture}
[
help/.style={circle,draw=white!100,fill=white!100,thick, inner sep=0pt,minimum size=1mm},
white/.style={circle,draw=black!100,fill=white!100,thick, inner sep=0pt,minimum size=2mm},
black/.style={circle,draw=black!100,fill=black!100,thick, inner sep=0pt,minimum size=2mm},
rect/.style={rectangle,draw=black!100,fill=black!20,thick, inner sep=0pt,minimum size=2mm},
wrect/.style={rectangle,draw=black!100,fill=white!100,thick, inner sep=0pt,minimum size=2mm},
brect/.style={rectangle,draw=black!100,fill=black!100,thick, inner sep=0pt,minimum size=2mm},
grey/.style={circle,draw=black!100,fill=black!20,thick, inner sep=0pt,minimum size=2mm},
xscale=0.86,
yscale=0.86
]


\node at (1,-0.2) [help](A1) [label=90: $\frm{A}_1$]{};

\node at (1,1) [white](rootA1) [ ]{};

\node at (.25,1.5) [white](rootA11) [ ]{};

\node at (1,2) [white](A11) [ ]{};

\node at (.4,2) [white](A11r) [ ]{};

\begin{scope}[>=stealth, auto]
\draw [->] (rootA1) to  (A11);

\draw [->] (rootA1) to  (A11r);

\draw [->] (rootA1) to  (rootA11);

\draw [->] (A11) to  (A11r);

\draw [->] (rootA1) [out=310, in=230, loop]to  (rootA1);

\draw [->] (A11) [out=50, in=130, loop]to  (A11);

\draw [->] (A11r) [out=50, in=130, loop]to  (A11r);

\draw [->] (rootA11) [out=70, in=150, loop]to  (rootA11);
\end{scope}

\node at (1.75,-0.2) [help](A2) [label=90: $\frm{A}_2$]{};

\node at (1.75,1) [white](rootA2) [ ]{};

\node at (1.75,2) [white](rootA21) [ ]{};

\begin{scope}[>=stealth, auto]
\draw [->] (rootA2) to  (rootA21);

\draw [->] (rootA2) [out=310, in=230, loop]to  (rootA2);

\draw [->] (rootA21) [out=50, in=130, loop]to  (rootA21);
\end{scope}

\node at (3,-0.2) [help](A3) [label=90: $\frm{A}_3$]{};

\node at (3,1) [white](rootA3) [ ]{};

\node at (3,2) [white](A31) [ ]{};

\node at (2.4,2) [white](rootA31) [ ]{};

\begin{scope}[>=stealth, auto]
\draw [->] (rootA3) to  (A31);

\draw [->] (rootA3) to  (rootA31);

\draw [->] (rootA3) [out=310, in=230, loop]to  (rootA3);

\draw [->] (rootA31) [out=50, in=130, loop]to  (rootA31);

\draw [->] (A31) [out=50, in=130, loop]to  (A31);
\end{scope}

\node at ( 3.25,2.75) [help](helpUpDotted) []{};
\node at ( 3.25,0) [help](helpDownDotted) []{};

\begin{scope}[>=stealth, auto]



\draw [ dotted] (helpUpDotted) to  (helpDownDotted);


\end{scope}

\node at (4.4,-0.2) [help](1B) [label=90: $\frm{B}_1$]{};
\node at (4.4,1) [white](1rootB) [ ]{};
\node at (4.4,2) [white](1B1) [ ]{};
\node at (3.8,2) [white](1B2) [ ]{};
\node at (3.65,1.5) [white](1rootB1) [ ]{};
\begin{scope}[>=stealth, auto]
\draw [->] (1rootB) to  (1B1);
\draw [->] (1B1) to  (1B2);
\draw [->] (1rootB) to  (1rootB1);

\draw [->] (1rootB) [out=310, in=230, loop]to  (1rootB);
\draw [->] (1B1) [out=50, in=130, loop]to  (1B1);
\draw [->] (1B2) [out=50, in=130, loop]to  (1B2);
\draw [->] (1rootB1) [out=70, in=150, loop]to  (1rootB1);

\end{scope}


\node at (5,-0.2) [help](2B) [label=90: $\frm{B}_2$]{};
\node at (5,1) [white](2rootB) [ ]{};
\node at (5,2) [white](2B1) [ ]{};

\begin{scope}[>=stealth, auto]
\draw [->] (2rootB) to  (2B1);
\draw [->] (2B1) [out=50, in=130, loop]to  (2B1);

\end{scope}



\node at (6.75,-0.3) [help](mA1) [label=90: $\mods{\frm A}_1$]{};

\node at (6.75,1) [wrect](mrootA1) [ label=180:$\triangleright$ ]{};

\node at (6,1.5) [black](mrootA11) [ ]{};

\node at (6.75,2) [white](mA11) [ ]{};

\node at (6.15,2) [rect](mA11r) [ ]{};

\begin{scope}[>=stealth, auto]
\draw [->] (mrootA1) to  (mA11);

\draw [->] (mrootA1) to  (mA11r);

\draw [->] (mrootA1) to  (mrootA11);

\draw [->] (mA11) to  (mA11r);

\draw [->] (mrootA1) [out=310, in=230, loop]to  (mrootA1);

\draw [->] (mA11) [out=50, in=130, loop]to  (mA11);

\draw [->] (mA11r) [out=50, in=130, loop]to  (mA11r);

\draw [->] (mrootA11) [out=70, in=150, loop]to  (mrootA11);
\end{scope}

\node at (7.7,-0.3) [help](mA2) [label=90: $\mods{\frm A}_2$]{};

\node at (7.7,1) [wrect](mrootA2) [ label=180:$\triangleright$ ]{};

\node at (7.7,2) [black](mrootA21) [ ]{};

\begin{scope}[>=stealth, auto]
\draw [->] (mrootA2) to  (mrootA21);

\draw [->] (mrootA2) [out=310, in=230, loop]to  (mrootA2);

\draw [->] (mrootA21) [out=50, in=130, loop]to  (mrootA21);
\end{scope}


\node at (8.85,-0.3) [help](mA3) [label=90: $\mods{\frm A}_3$]{};

\node at (8.85,1) [wrect](mrootA3) [ label=180:$\triangleright$ ]{};

\node at (8.85,2) [white](mA31) [ ]{};

\node at (8.25,2) [black](mrootA31) [ ]{};

\begin{scope}[>=stealth, auto]
\draw [->] (mrootA3) to  (mA31);

\draw [->] (mrootA3) to  (mrootA31);

\draw [->] (mrootA3) [out=310, in=230, loop]to  (mrootA3);

\draw [->] (mrootA31) [out=50, in=130, loop]to  (mrootA31);

\draw [->] (mA31) [out=50, in=130, loop]to  (mA31);
\end{scope}

\node at (9.75,-0.3) [help](mA4) [label=90: $\mods{\frm A}_4$]{};

\node at (9.75,1) [wrect](mrootA4) [ label=180:$\triangleright$]{};

\node at (9.75,2) [brect](mrootA41) [ ]{};

\begin{scope}[>=stealth, auto]
\draw [->] (mrootA4) to  (mrootA41);

\draw [->] (mrootA4) [out=310, in=230, loop]to  (mrootA4);

\draw [->] (mrootA41) [out=50, in=130, loop]to  (mrootA41);
\end{scope}


\node at (10.6,-0.3) [help](mA5) [label=90: $\mods{\frm A}_5$]{};

\node at (10.6,1) [wrect](mrootA5) []{};

\node at (10.6,2) [black](mrootA51) [ label=180:$\triangleright$]{};

\begin{scope}[>=stealth, auto]
\draw [->] (mrootA5) to  (mrootA51);

\draw [->] (mrootA5) [out=310, in=230, loop]to  (mrootA5);

\draw [->] (mrootA51) [out=50, in=130, loop]to  (mrootA51);
\end{scope}

\node at (10.9,2.75) [help](mhelpUpDotted) []{};
\node at ( 10.9,0) [help](mhelpDownDotted) []{};

\begin{scope}[>=stealth, auto]



\draw [ dotted] (mhelpUpDotted) to  (mhelpDownDotted);


\end{scope}



\node at (12.1,-0.3) [help](m1B) [label=90: $\mods{\frm B}_1$]{};
\node at (12.1,1) [wrect](m1rootB) [ label=180:$\triangleright$ ]{};
\node at (12.1,2) [white](m1B1) [ ]{};
\node at (11.5,2) [rect](m1B2) [ ]{};
\node at (11.26,1.5) [black](m1rootB1) [ ]{};
\begin{scope}[>=stealth, auto]
\draw [->] (m1rootB) to  (m1B1);
\draw [->] (m1B1) to  (m1B2);
\draw [->] (m1rootB) to  (m1rootB1);

\draw [->] (m1rootB) [out=310, in=230, loop]to  (m1rootB);
\draw [->] (m1B1) [out=50, in=130, loop]to  (m1B1);
\draw [->] (m1B2) [out=50, in=130, loop]to  (m1B2);
\draw [->] (m1rootB1) [out=70, in=150, loop]to  (m1rootB1);

\end{scope}


\node at (13,-0.3) [help](m2B) [label=90: $\mods{\frm B}_2$]{};
\node at (13,1) [grey](m2rootB) [ label=180:$\triangleright$ ]{};
\node at (13,2) [brect](m2B1) [ ]{};

\begin{scope}[>=stealth, auto]
\draw [->] (m2rootB) to  (m2B1);
\draw [->] (m2B1) [out=50, in=130, loop]to  (m2B1);

\end{scope}

\node at ( 5.5,-0.2) [help](mhelpDown) []{};
\node at ( 5.5,3) [help](mhelpUp) []{};
\node at ( -0.1,3) [help](mhelpleftUp) []{};
\node at ( -0.1,-0.2) [help](mhelpleftDown) []{};
\node at ( 13.5,3) [help](mhelpRightUp) []{};
\node at ( 13.5,-0.2) [help](mhelpRightDown) []{};

\begin{scope}[>=stealth, auto]

\draw  (mhelpUp) to  (mhelpDown);
\draw  (mhelpleftUp) to  (mhelpleftDown);

\draw  (mhelpRightUp) to  (mhelpRightDown);

\draw  (mhelpUp) to  (mhelpRightUp);
\draw  (mhelpUp) to  (mhelpleftUp);
\draw   (mhelpDown) to (mhelpleftDown);
\draw   (mhelpDown) to (mhelpRightDown);
\end{scope}

\end{tikzpicture}
\end{center}
\caption{The frames  $\frm{A}_1$, $\frm{A}_2$, $\frm{A}_3$ and $\frm B_1$, $\frm{B}_2$, and the pointed models based on them.}
\label{fig:s4}
\end{figure}
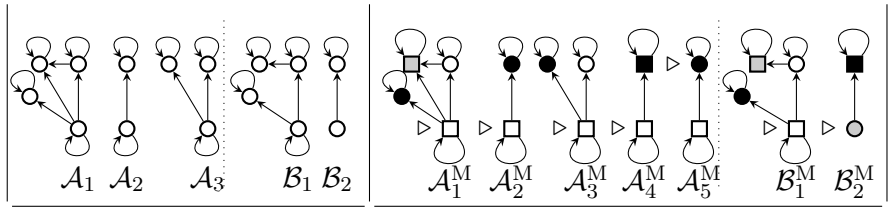

Let us consider a  $( \lang_{\pd},\langle \cls{A},\cls{B}\rangle)$-{\fsgf} where
$\cls{A}=\{\frm{A}_1, \frm{A}_2, \frm{A}_3\}$ and $\cls{B}=\{\frm{B}_1,\frm{B}_2\}$ as  shown in the left rectangle  in Figure~\ref{fig:s4}.
Obviously, $(\overline{p} \wedge \Box\Box\overline{p})\vee\pd p$ is valid on all the frames in $\cls{A}$ and not valid on any frame in $\cls{B}$.
As before, Hercules and the Hydra will play a frame equivalence game on $\langle \cls A,\cls B\rangle$.\\

\noindent{\sc selection of the models on the right:} Hercules must choose his models as  follows.

\begin{lemma}\label{lem:s4A1B1A2B2}
In any winning strategy for Hercules for an $( \lang_{\pd},\langle \cls{L},\cls{R}\rangle)$-{\fsgf}, 
\begin{enumerate}

\item
if $\frm{A}_1\in \cls{L}$ and $\frm{B}_1\in \cls{R}$, Hercules picks a pointed model $(\mods{\frm B}_1,\point)$ based on the lowest point in $\frm{B}_1$ and
 assigns different valuations to the  two  leftmost  points;

\item if $\frm{A}_2\in \cls{L}$ and $\frm{B}_2\in \cls{R}$, Hercules picks a pointed model $(\mods{\frm B}_2,\point)$ based on the lowest point in $\frm{B}_2$ and
 assigns different valuations to the   points of $\frm{B}_2$.

\end{enumerate}

\end{lemma}

\proof
The proof of the first item is the same as the proof of Lemma~\ref{lemA1BTrans}. For the second item, 
if Hercules picked a pointed model based on the reflexive point in $\frm{B}_2$ or assigned the same valuations to the two points in $\frm{B}_2$,
 the Hydra would reply with a bisimilar pointed model based 
on $\frm{A}_2$ by making both points in $\frm{A}_2$ satisfy the same literals as the ones satisfied by the reflexive point in $\frm{B}_2$.
\endproof

In Figure \ref{fig:s4} we have indicated each point of $\frm B_1$ or $\frm B_2$ with a different shape or colour, but this is only meant to help visualize the Hydra's strategy; points with different shapes do not necessarily receive different valuations. The exception are the two leftmost points of ${\frm B}_1$, whose valuations must be different from each other's, as well as the two points of $\frm{B}_2$.
Thus, Hercules
constructs the pointed model $(\mods{\frm{B}}_1, \triangleright)$ and $(\mods{\frm{B}}_2, \triangleright)$ shown in the right rectangle in Figure~\ref{fig:s4}.
\medskip

\noindent{\sc selection of models on the left:}
The Hydra replies  by mimicking Hercules' choice as shown  on the left of the dotted line in the right rectangle in Figure~\ref{fig:s4}.
Points with the same shape and colour satisfy the same literals.
\medskip

\noindent{\sc formula size game on models:}
We consider the {\fsgm} starting with  
$(\mods{\frm A}_1, \triangleright),\ldots, ( \mods {\frm A}_5, \triangleright)$ on the left and $(\mods{\frm B}_1, \triangleright), (\mods{\frm B}_2, \triangleright)$
on the right. 

\begin{lemma}\label{lemTransNoModB}
Suppose that $\cls L$ and $\cls R$ are classes of models and Hercules has a winning strategy for the $( \lang_{\pd},\langle \cls{L},\cls{R}\rangle)$-{\fsgm}.
Suppose that $T$ is a closed game tree and the Hydra played greedily.
\begin{enumerate}

\item\label{itTransNoModOne} If $(\mods{\frm A}_1, \triangleright)\in \cls{L}$  and $(\mods{\frm B}_1, \triangleright)\in \cls{R}$, then there is a node
$\eta$ in $T$ with $(\mods{\frm A}_1, \triangleright)\in \lft(\eta)$  and $(\mods{\frm B}_1, \triangleright)\in \rgt(\eta)$ such that Hercules played a $\pd$-move at $\eta$.

\item\label{itTransNoModTwo} If  $\eta$ is a node of $T$ with $(\mods{\frm A}_5,\triangleright) \in \lft(\eta)$ and $(\mods{\frm B}_1, \triangleright) \in \rgt(\eta)$,
 then Hercules did not play a $\pd$-move at $\eta$.

\item\label{itTransNoModThree} If  $\eta$ is a node of $T$ with $(\mods{\frm A}_4,\triangleright) \in \lft(\eta)$ and there is a pointed model based on $\mods{\frm B}_2$ in  $\rgt(\eta)$,
 then Hercules did not play a $\Box$-move at $\eta$.

\end{enumerate}
\end{lemma}

\proof\

\noindent\eqref{itTransNoModOne}
Since  $(\mods{\frm A}_1, \triangleright)$ and $(\mods{\frm B}_1, \triangleright)$ satisfy the same literals, Hercules 
cannot play a literal move at a node $\chi$ with $(\mods{\frm A}_1, \triangleright)\in \lft(\chi)$  and $(\mods{\frm B}_1, \triangleright)\in \rgt(\chi)$.
 Playing a $\vee$- or a $\wedge$-move at such a node $\chi$  
 will result in at least one new game position $\kappa$ such that  $(\mods{\frm A}_1, \triangleright)\in \lft(\kappa)$  and $(\mods{\frm B}_1, \triangleright)\in \rgt(\kappa)$.
If Hercules played a $\Box$-move at $\chi$, he must have picked   again $(\mods{\frm B}_1, \triangleright)$ as a successor of $(\mods{\frm B}_1, \triangleright)$
because if he selected a pointed model based on either the black or the white point, the Hydra would reply with the same point in $\mods{\frm A}_1$ which
contradicts the fact that $T$ is closed. Therefore, playing a $\Box$-move at $\chi$ would result in a game position $\gamma$ with $(\mods{\frm A}_1, \triangleright)\in \lft(\gamma)$  and $(\mods{\frm B}_1, \triangleright)\in \rgt(\gamma)$. Thus, $T$ must contain a node $\eta$  with $(\mods{\frm A}_1, \triangleright)\in \lft(\eta)$  and $(\mods{\frm B}_1, \triangleright)\in \rgt(\eta)$ at which Hercules played a $\pd$-move.\\

\noindent\eqref{itTransNoModTwo} If Hercules played such a move he must select again $(\mods{\frm A}_5,\triangleright)$  as a successor of $(\mods{\frm A}_5,\triangleright)$
and the Hydra would reply with, among others, a bisimilar pointed model based on the black point in  $\mods{\frm B}_1$.\\

\noindent\eqref{itTransNoModThree} If Hercules played a $\Box$-move he must pick  a pointed model based on the black rectangle point in $\mods{\frm B}_1$
to which the Hydra would reply with the bisimilar successor of $(\mods{\frm A}_5,\triangleright)$.
\endproof

The next two lemmas establish how the game progresses as Hercules chooses new pointed models.

\begin{lemma}\label{lemTransNoModTwo}
Suppose that $\cls L$ and $\cls R$ are classes of models and Hercules has a winning strategy for the $( \lang_{\pd},\langle \cls{L},\cls{R}\rangle)$-{\fsgm}.
Suppose that $T$ is a closed game tree and the Hydra played greedily.
\begin{enumerate}

\item\label{itTransNoModFour} If  $\eta$ is a node of $T$ with $(\mods{\frm A}_4,\triangleright) \in \lft(\eta)$ and there is a pointed model based on $\mods{\frm B}_2$ in  $\rgt(\eta)$,
 then if Hercules played a $\pd$-move at $\eta$, he selected again $(\mods{\frm A}_4,\triangleright)$.

\item\label{itTransNoModFive}  Let $\eta$ be a node on which Hercules played a $\Box$-move. If, for some $i\in \{2,3\}$, $(\mods{\frm{A}}_i, \triangleright) \in \lft(\eta)$ and $(\mods{\frm{B}}_1, \triangleright)\in \rgt(\eta)$, then he did not pick a pointed model based on the black point  as a successor of $(\mods{\frm B }_1, \triangleright)$.

\item\label{itTransNoModSix}  Let $\eta$ be a node on which Hercules played a $\pd$-move. If, for some $i\in \{2,3\}$, $(\mods{\frm{A}}_i, \triangleright) \in \lft(\eta)$ and $(\mods{\frm{B}}_1, \triangleright)\in \rgt(\eta)$, then he did not pick a pointed model based on the black point  as a successor of $(\mods{\frm A }_i, \triangleright)$.

\end{enumerate}
\end{lemma}

\proof\

\noindent\eqref{itTransNoModFour} If Hercules picked a pointed model based on the black rectangle point in $\mods{\frm A}_4$,
the Hydra would reply with the same point in $\mods{\frm B}_2$.
\\

\noindent \eqref{itTransNoModFive}-\eqref{itTransNoModSix} If  Hercules picked a pointed model based on the black point the Hydra would reply with the same point in
$\mods{\frm B}_1$.
In either case we obtain bisimilar models on each side, which by Corollary~\ref{lemHercLose} means that Hercules cannot win.
\endproof

\begin{lemma}\label{lem:boxmovesB}
Let  $\cls L$ and $\cls R$ be classes of pointed models for which Hercules has a winning strategy in the $( \lang_{\pd},\langle \cls{L},\cls{R}\rangle)$-{\fsgm}.
Let $T$ be a closed game tree for this game  and let us suppose that the  Hydra played greedily. Let $(\mods{\frm A}_3, w)$ and $(\mods{\frm B}_1,w)$ 
denote the pointed models based on the respective model and the white circular point in it.
\begin{enumerate}
\item \label{itBoxmovesOne} If $(\mods{\frm A}_2,\triangleright)\in \cls{L}$ and $(\mods{\frm B}_1,\triangleright)\in \cls{R}$, then there is 
a node $\eta$ in $T$ such that  $(\mods{\frm A}_2,\triangleright)\in\lft(\eta)$, $(\mods{\frm B}_1,\triangleright)\in\rgt(\eta)$,
and Hercules played a $\Box$-move at $\eta$ so that he selected  $(\mods{\frm B}_1,w)$.
\item \label{itBoxmovesTwo} If $(\mods{\frm A}_3,\triangleright)\in \cls{L}$ and $(\mods{\frm B}_1,\triangleright)\in \cls{R}$, then there is 
a node $\eta$ in $T$ such that  $(\mods{\frm A}_3,w)\in\lft(\eta)$, $(\mods{\frm B}_1,w)\in\rgt(\eta)$.
\item \label{itBoxmovesThree} If $(\mods{\frm A}_3,w)\in \cls{L}$ and $(\mods{\frm B}_1,w)\in \cls{R}$, then there is 
a node $\eta$ in $T$ such that  $(\mods{\frm A}_3,w)\in\lft(\eta)$, $(\mods{\frm B}_1,w)\in\rgt(\eta)$
and Hercules played a $\Box$-move at $\eta$.
\end{enumerate}
\end{lemma}

\proof\

\noindent\eqref{itBoxmovesOne} Since $(\mods{\frm A}_2,\triangleright)$ and $(\mods{\frm B}_1,\triangleright)$ satisfy the same literals, 
Hercules cannot play a literal move at a node $\eta$ with  $(\mods{\frm A}_2,\triangleright)\in \lft(\eta)$ and $(\mods{\frm B}_1,\triangleright)\in \rgt(\eta)$.
Playing a $\vee$- or a $\wedge$-move would result in at least one new tree-node $\chi$ with 
$(\mods{\frm A}_2,\triangleright)\in \lft(\chi)$ and $(\mods{\frm B}_1,\triangleright)\in \rgt(\chi)$. Using the last item 
of Lemma~\ref{lemTransNoModTwo}, we see that, if Hercules plays a $\pd$-move at such a node, he is going to
pick  $(\mods{\frm A}_2,\triangleright)$ again to which the Hydra is going to reply with, among others, $(\mods{\frm B}_1,\triangleright)$
and thus we are back in essentially the same game position. The same is true if Hercules plays a $\Box$-move and selects $(\mods{\frm B}_1,\triangleright)$.
Therefore, using the second item of Lemma~\ref{lemTransNoModTwo}, we conclude that there must be a node $\eta$ in $T$ such that  
$(\mods{\frm A}_2,\triangleright)\in\lft(\eta)$, $(\mods{\frm B}_1,\triangleright)\in\rgt(\eta)$,
and Hercules played a $\Box$-move at $\eta$ so that he selected  $(\mods{\frm B}_1,w)$.\\

\noindent\eqref{itBoxmovesTwo} The proof of this item is almost immediate with the help of the last two items of Lemma~\ref{lemTransNoModTwo}.\\

\noindent\eqref{itBoxmovesThree}
 Since $(\mods{\frm A}_3,w)$ and $(\mods{\frm B}_1,w)$ satisfy the same literals, 
Hercules cannot play a literal move at a node $\chi$ with $(\mods{\frm A}_3,w)\in\lft(\chi)$ and $(\mods{\frm B}_1,w)\in\rgt(\chi)$.
Playing a $\vee$- or a $\wedge$-move would result in at least one new tree-node $\kappa$ with 
$(\mods{\frm A}_2,w)\in \lft(\kappa)$ and $(\mods{\frm B}_1,w)\in \rgt(\kappa)$. Obviously,
 if Hercules plays a $\pd$-move at such a node, he is going to
pick  $(\mods{\frm A}_3,w)$ again to which the Hydra is going to reply with, among others, $(\mods{\frm B}_1,w)$
and thus we are back in essentially the same game position. Hence,   there must be
a node $\eta$ in $T$ such that  $(\mods{\frm A}_3,w)\in\lft(\eta)$, $(\mods{\frm B}_1,w)\in\rgt(\eta)$,
and Hercules played a $\Box$-move at $\eta$.
\endproof

With this we are ready to prove Theorem \ref{s4axiom}. The next proposition is essentially a more explicit version of the theorem.

\begin{proposition}\label{propTransBoundsB}
Let $\cls L$ and $\cls R$ be classes of models such that Hercules has a winning strategy for the $( \lang_{\pd},\langle \cls{L},\cls{R}\rangle)$-{\fsgm}.
Let $T$ be a closed game tree in which the Hydra played greedily.
\begin{enumerate}
\item \label{itTransDiamB} If $(\mods{\frm{A}}_1, \triangleright)\in \cls{L}$ and $(\mods{\frm B}_1, \triangleright)\in \cls{R}$, 
then Hercules  has made at least one $\pd$-move.

\item\label{itTransDisB} If $(\mods{\frm A}_1, \triangleright), (\mods{\frm A}_5, \triangleright) \in \cls{L}$ and $(\mods{\frm B}_1, \triangleright)\in \cls{R}$, then Hercules made
at least one $\vee$-move during the game.

\item\label{itTransConj} If $(\mods{\frm A}_4, \triangleright) \in \cls{L}$ and $(\mods{\frm B}_1, \triangleright), (\mods{\frm B}_2, \triangleright)\in \cls{R}$, 
then Hercules made
at least one $\wedge$-move during the game.

\item\label{itTransBoxB}
If $\{(\mods{\frm A}_2, \triangleright), (\mods{\frm A}_3,\triangleright)\}\subseteq \cls{L}$ and $(\mods{\frm B}_1, \triangleright)\in \cls{R}$,
 then Hercules played  at least two $\Box$-moves.

\end{enumerate}

\end{proposition}

\proof
The first and the second item follow from the first two items of  Lemma \ref{lemTransNoModB}. 
For the third item, let us suppose that Hercules did not play an $\wedge$-move. Since, 
$(\mods{\frm A}_4, \triangleright)$ and $(\mods{\frm B}_1, \triangleright)$ satisfy the same literals, Hercules 
cannot play a literal move at a node $\chi$ with $(\mods{\frm A}_4, \triangleright)\in \lft(\chi)$  while $(\mods{\frm B}_1, \triangleright)$ and
a pointed model $\frm{M}$ based on  $\mods{\frm B}_2$ are in  $\rgt(\chi)$.
 Playing a $\vee$-move at such a node $\chi$  
 will result in at least one new game position $\kappa$ such that  $(\mods{\frm A}_4, \triangleright)\in \lft(\kappa)$  and $(\mods{\frm B}_1, \triangleright)$ and  $\frm{M}$
are in $\rgt(\kappa)$. According to the third item of Lemma~\ref{lemTransNoModB}, Hercules is not going to play a $\Box$-move at such a node
whereas according to the first item of Lemma~\ref{lemTransNoModTwo}, if Hercules plays a $\pd$-move, he must select $(\mods{\frm A}_4, \triangleright)$,
to which the Hydra is going to reply with among others $(\mods{\frm B}_1, \triangleright)$ and a pointed model based on $\mods{\frm{B}}_2$ and we are
back in the previous situation.
Thus in the absence of a $\wedge$-move we see that the game-tree $T$ cannot be closed. 

For the last item, it follows from Lemma~\ref{lem:boxmovesB} that 
\begin{enumerate}[label=(\alph*)]

\item
 there is 
a node $\eta$ in $T$ such that  $(\mods{\frm A}_2,\triangleright) \in \lft ( \eta )$, $(\mods{\frm B}_1,\triangleright)\in\rgt(\eta)$,
and Hercules played a $\Box$-move at $\eta$ so that he selected  $(\mods{\frm B}_1,w)$;
\item 
there is a node $\chi$ in $T$ such that  $(\mods{\frm A}_3,w)\in\lft(\chi)$, $(\mods{\frm B}_1,w)\in\rgt(\chi)$
and Hercules played a $\Box$-move at $\chi$.

\end{enumerate}
If $\eta $ and $\chi$ do not coincide, then it is obvious that Hercules played at least two $\Box$-moves.
Let us suppose now that $\eta$ and $\chi$ coincide and let $\kappa$ be the successor node in $T$. According to the first item,
$(\mods{\frm B}_1,w)\in \rgt(\kappa)$. Using the fact that the Hydra plays greedily and the second item, we see that 
$(\mods{\frm A}_3,w)\in\lft(\kappa)$. It is immediate from the third item of Lemma~\ref{lem:boxmovesB} that 
in the sub-game starting at the node $\kappa$, Hercules played at least one additional $\Box$-move.  
\endproof
\noindent With this we conclude the proof of Theorem~\ref{s4axiom}.

\section{The L\"{o}b axiom}\label{secLob}

The L\"ob axiom defines the property of transitivity and converse-well-foundedness (i.e., that there are no infinite chains $w_0 \mathrel R w_1 \mathrel R \ldots$). This is a conjunction of two properties and the resulting {\em G\"odel-L\"ob logic} $\mathsf{GL}$ is often presented with the additional axiom $\Box p \to \Box \Box p $, but it is a non-trivial exercise to show that this is already a consequence of the L\"ob axiom $\Box(\Box p \to p)\to \Box p$. Note that well-foundedness is a second-order property, and cannot be defined in first-order logic.

\begin{theorem}\label{theoLob}
The formula $\Box\overline{p}\vee\pd(p\wedge\Box\overline{p})$ is absolutely minimal among all formulas defining the class of transitive and converse well-founded frames.
\end{theorem}
We have already shown that $\Box\Box \overline p \vee \pd p$ is absolutely minimal among those formulas defining transitivity, so our strategy will be to expand on the frames and pointed models in Figure \ref{fig:nsmallerthanm} to additionally force Hercules to play a conjunction.
Since these models were already well-founded we can use previous results.

\begin{figure} 
\begin{center}
\begin{tikzpicture}
[
help/.style={circle,draw=white!100,fill=white!100,thick, inner sep=0pt,minimum size=1mm},
white/.style={circle,draw=black!100,fill=white!100,thick, inner sep=0pt,minimum size=2mm},
black/.style={circle,draw=black!100,fill=black!100,thick, inner sep=0pt,minimum size=2mm},
xscale=0.9,
yscale=0.9
]


\node at (1,1.1) [help](A1) [label=90: $\frm{A}_1$]{};
\node at (1,2) [white](rootA1) [ ]{};
\node at (0.4,2.5) [white](rootA11) [ ]{};
\node at (1,3) [white](A11) [ ]{};
\node at (0.4,3) [white](A11r) [ ]{};

\begin{scope}[>=stealth, auto]
\draw [->] (rootA1) to  (A11);
\draw [->] (rootA1) to  (A11r);
\draw [->] (rootA1) to  (rootA11);
\draw [->] (A11) to  (A11r);

\end{scope}

\node at (2,1.1) [help](A2) [label=90: $\frm{A}_2$]{};
\node at (2,2) [white](rootA2) [ ]{};
\node at (1.4,3) [white](rootA21) [ ]{};

\begin{scope}[>=stealth, auto]
\draw [->] (rootA2) to  (rootA21);
\end{scope}

\node at (3,1.1) [help](A3) [label=90: $\frm{A}_3$]{};
\node at (3,2) [white](rootA3) [ ]{};
\node at (3,3) [white](A31) [ ]{};
\node at (2.4,3) [white](rootA31) [ ]{};

\begin{scope}[>=stealth, auto]
\draw [->] (rootA3) to  (A31);
\draw [->] (rootA3) to  (rootA31);
\end{scope}

\node at ( 4,2) [white](rootA) [ ]{};
\node at ( 4,3) [white](a11) [ ]{};
\node at ( 5,2.5) [white](a21) [ ]{};
\node at ( 6,3) [white](a22) [ ]{};

\node at ( 5,2) [white](a31) [ ]{};
\node at ( 6,2) [white](a32) [ ]{};
\node at ( 7,2) [white](a33) [ ]{};

\node at (4,1.1) [help](A) [label=90: $\mathcal{A}_4$]{};
\node at (5.5,1.5) [help](Adots) []{{\bf $\vdots$}};

\begin{scope}[>=stealth, auto]

\draw [->] (rootA) to  (a11);
\draw [->] (rootA) to  (a21);
\draw [->] (rootA) to  (a31);
\draw [->] (a21) to  (a22);

\draw [->] (a31) to  (a32);
\draw [->] (a32) to  (a33);

\end{scope}

\node at ( 7.5,3.3) [help](helpUpDotted) []{};
\node at ( 7.5,1) [help](helpDownDotted) []{};

\begin{scope}[>=stealth, auto]



\draw [ dotted] (helpUpDotted) to  (helpDownDotted);


\end{scope}
\node at (8.8,1.1) [help](oB) [label=90: $\frm{B}$]{};
\node at (8.8,2) [white](orootB) [ ]{};
\node at (8.8,3) [white](oB1) [ ]{};

\node at (8,3) [white](oB4) [ ]{};
\node at (8,2.6) [white](orootB1) [ ]{};

\begin{scope}[>=stealth, auto]
\draw [->] (orootB) to  (oB1);
\draw [->] (orootB) to  (orootB1);
\draw [->] (oB1) to  (oB4);
\end{scope}

\node at (9.8,2) [white](rootB1) [ ]{};
\node at (10.8,2) [white](deadend) [ ]{};
\node at (9.8,1.05) [help](B1) [label=90: $\mathcal{B}_1$]{};

\begin{scope}[>=stealth, auto]

\draw [->] (rootB1) to  (deadend);
\draw [->] (rootB1) [out=50, in=130, loop]to  (rootB1);
\end{scope}

\node at ( 0,3.5) [help](mhelpleftUp) []{};
\node at ( 0,.8) [help](mhelpleftDown) []{};
\node at ( 11.25,3.5) [help](mhelpRightUp) []{};
\node at ( 11.25,.8) [help](mhelpRightDown) []{};
\begin{scope}[>=stealth, auto]
\draw  (mhelpleftUp) to  (mhelpleftDown);
\draw  (mhelpRightUp) to  (mhelpRightDown);
\draw   (mhelpRightDown) to (mhelpleftDown);
\draw   (mhelpRightUp) to (mhelpleftUp);
\end{scope}

\end{tikzpicture}
\end{center}
\caption{The sets of  frames  $\cls{A}=\{{\frm A}_{1}, {\frm A}_{2}, {\frm A}_{3}, {\frm A}_{4}\}$ and $\cls{B}=\{{\frm B}, {\frm B}_{1}\}$.}
\label{fig:lobAxiom}
\end{figure}
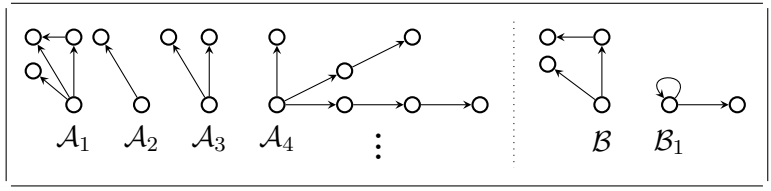

Let us consider an $( \lang_{\pd},\langle \cls{A},\cls{B}\rangle)$-{\fsgf} played by Hercules and the Hydra
with the frames shown  in Figure~\ref{fig:lobAxiom}. Obviously,
$\frm{A}_1, \frm{A}_2, \frm{A}_3$, and 
$\frm{B}$ are obtained from the frames  in Figure \ref{fig:nsmallerthanm} for $m=2$ and $n=1$.
Additionally, $\cls{A}$ contains the frame $\frm A_4$ that is
a transitive tree with infinitely many branches such that, for every natural number $n>0$, there is a branch
 for which the maximum number of relation steps from the root to its leaf is $n$. Similarly, $\cls{B}$ contains 
the frame $\frm{B}_1$ shown on the right of the dotted line in the same figure. Intuitively, we are
going to use ${\frm A}_{4}$ and ${\frm B}_{1}$  in order to force Hercules to play an $\wedge$-move.\\

\noindent{\sc selection of the models on the right:}
We only consider the choice of pointed model for the frame $\frm{B}_1$. It is obvious that Hercules is not going to base a pointed model on the dead-end point in $\frm{B}_1$ because the Hydra
would reply with a bisimilar pointed model based on one of the leaves of $\frm{A}_4$.

\begin{lemma}\label{lemLobPoint}
In any winning strategy for Hercules in the $(\lang_\pd,\langle \cls A,\cls B \rangle)$-{\fsgf}, Hercules will choose a pointed model based on the reflexive point on $\frm B_1$.
\end{lemma}

\noindent{\sc selection of models on the left:}
Hydra will choose her pointed models based on $\frm A_1$, $\frm A_2$ and $\frm A_3$ as before. For her pointed model based on $\frm{A}_4$, she picks a pointed model based on the root of the tree in which
 all leaves of $\frm{A}_4$ satisfy the same literals as the ones satisfied by the dead-end point
in $\frm{B}_1$ whereas the rest of the points satisfy the same literals as the ones satisfied by the reflexive point in $\frm{B}_1$. Once again, if Hydra plays in this way we say that she {\em mimics} Hercules' selection.
\medskip

\noindent{\sc formula size game on models:} 
The next lemmas will be used to prove that Hercules must play an $\wedge$-move.

\begin{lemma}\label{lemLobAgain}
Let $\cls L$, $\cls R$ be classes of models such that Hercules has a winning strategy for the $( \lang_{\pd},\langle \cls{L},\cls{R}\rangle)$-{\fsgm}.
If $T$ is a closed game tree on which the Hydra played greedily, then for any game position $\eta$ and any non-leaf point $w$ of $\frm{A}_4$, if $(\mods{\frm A} _4, w) \in \lft(\eta)$, $(\mods{\frm B}_1,\triangleright)\in \rgt(\eta)$, and Hercules played a $\Box$-move at $\eta$, then he selected $(\mods{\frm B}_1, \triangleright)$ again.

\end{lemma}

\proof
If Hercules picked the dead-end point in $\mods{\frm B}_1$, the Hydra, using the transitivity of the relation, would reply with a bisimilar pointed model
based on a leaf in $\mods{\frm A}_4$.
\endproof

\begin{proposition}\label{propLobDia}
Suppose that $\cls L$, $\cls R$ are classes of models for which Hercules has a winning strategy for the $( \lang_{\pd},\langle \cls{L},\cls{R}\rangle)$-{\fsgm} and let $T$ be a closed game tree on which the Hydra played greedily.
\begin{enumerate}

\item\label{itLobOne} If $(\mods{\frm A}_4, \triangleright)\in \cls{L}$ and $(\mods{\frm B}_1, \triangleright)\in \cls{R}$, Hercules played at least one $\pd$-move on a node $\eta$ such that 
$\lft(\eta)$ contains a pointed model based on $\mods{\frm A}_4$ whereas $(\mods{\frm B}_1,\triangleright) \in \rgt(\eta)$.

\item \label{itLobTwo} If Hercules  plays a $\pd$-move in a position $\eta$ in which $\lft(\eta)$ contains a pointed model  based on $\mods{\frm A}_4$  while $(\mods{\frm B}_1,\triangleright)$ is on the right, he must play at least one subsequent $\wedge$-move.

\end{enumerate}
\end{proposition}

\proof\

\noindent\eqref{itLobOne} Let us suppose that Hercules plays without $\pd$-moves. Since $(\mods{\frm A}_4,\triangleright)$ and $(\mods{\frm B}_1,\triangleright)$
satisfy the same literals, no literal move is possible in a game position $\eta$ in which $(\mods{\frm A}_4,\triangleright)$ is on the left and $(\mods{\frm B}_1,\triangleright)$
on the right. Playing a $\wedge$- or a $\vee$-move results in at least one new position in which $(\mods{\frm A}_4,\triangleright)$ is on the left and $(\mods{\frm B}_1,\triangleright)$
is on the right. Hence a $\Box$-move is inevitable and by Lemma \ref{lemLobAgain}, he selected $(\mods{\frm B}_1, \triangleright)$ again.

When Hercules plays such a move, the Hydra replies with all infinitely many pointed models based on an immediate successor of the root of $\mods{\frm A}_4$. From this new position on any finite number of $\vee$-, $\wedge$- and $\Box$-moves
are going to result in at least one new position that contains $(\mods{\frm B}_1,\triangleright)$ on the right
whereas on the left we have infinitely many pointed models based on $\mods{\frm A}_4$ and a non-leaf point. Obviously, none of the $\top$-, $\bot$-, and literal-moves are
possible in such a position. Hence, Hercules has no winning strategy without $\pd$-moves. 
\smallskip

\noindent \eqref{itLobTwo} 
Let us suppose that Hercules plays a $\pd$-move in such a position. 
The Hydra is going to respond with both $(\mods{\frm B}_1,\triangleright)$ and a pointed model based on the dead-end point in $\mods{\frm B}_1$.
Let us suppose now that Hercules is not going to play any subsequent $\wedge$-move. Obviously, $\bot$, $\top$, and literal moves are impossible; moreover,
the presence of a dead-end pointed model on the right prevents $\Box$-moves. Clearly, playing an $\vee$-move
would result in at least one new game position which is the same as the previous one. Therefore, Hercules can only play $\pd$-moves until he reaches 
a pointed model $(\frm{A}_4, v)$ such that the only successor of $v$ is a leaf. Playing a $\pd$-move in such a position would lead to a loss in the next step because of the presence
of bisimilar pointed models on the left and right. Since $(\mods{\frm A}_4, v)$ and $(\mods{\frm B}_1,\triangleright)$ satisfy the same literals
no literal moves are possible either. Therefore, Hercules has no winning strategy without playing at least one $\wedge$-move.
\endproof

\noindent With this we can prove Theorem \ref{theoLob}.

\proof
Consider a $(\lang_\pd,\langle \cls A,\cls B\rangle)$-{\fsgf} where $\cls A = \{\frm A_1,\frm A_2,\frm A_3,\frm A_4\}$ and $\cls B= \{ \frm B,\frm B_1 \}$ as given in Figure 
\ref{fig:lobAxiom}. Hercules must choose his pointed models according to Lemmas \ref{lemA1BTrans} and \ref{lemLobPoint}, and Hydra replies by mimicking Hercules. Using Proposition \ref{propTransBoundsA} we see that if the Hydra plays greedily then any closed game tree must have modal depth at least two, contain two instances of $\Box$, one instance of each $\pd$ and $\vee$, and one variable. By Proposition \ref{propLobDia}, it also contains one conjunction, as required.
\endproof

\section{Conclusion}\label{secConclusion}
The present work was  motivated to a large degree by  ideas and results from \cite{vakarelov}, where 
the notion of minimal modal equivalent of a first-order condition was introduced.
Note however that the term `minimal' is used in  \cite{vakarelov} only with respect to the 
number of different variables needed to modally define a first-order condition:
this does not tell us much about the length, modal depth, or the number of Boolean connectives required and that is why we have 
extended the notion of minimality to cover these as well. With this we have provided lower bounds on non-colourability axioms and shown that several familiar modal axioms are minimal with respect to all measures considered, including the L\"ob axiom.
Note that neither non-colourability nor the L\"ob axiom are first-order definable.

It is obvious that once we have shown that a given frame property is modally definable,
we can study its minimal modal complexity with respect to different complexity measures  
and therefore there are many natural open problems related to the
present work. We would like to mention one in particular.
The importance of  the Sahlqvist formulae cannot be overstated and they have been studied 
extensively  over  the  years. However, it seems that a very basic question about them has not received 
the attention it deserves. Namely, since these formulae  have a specific ``syntactic shape'', it is natural
to ask whether this syntactic restriction   leads to an increase of their complexity. 
It was conjectured in \cite{vakarelov} that there are first-order conditions that
can be defined by both non-Sahlqvist and Sahlqvist formulae but the latter 
require more propositional variables than the former. 
This conjecture is an instance of the following general problem

\begin{question}
Is there a complexity measure $\mu$  with respect to which  Sahlqvist formulae are asymptotically  more complex 
than non-Sahlqvist ones and by how much? In particular, can this complexity gap be ``big'', i.e,  
is there a natural complexity measure $\mu$ and an infinite sequence of formulae  $\varphi_1,\varphi_2,\ldots$ such that
if $\psi_1,\psi_2,\ldots $ is a sequence of equivalent Sahlqvist formulae then $\mu(\psi_n)$ grows 
super-polynomially or even exponentially in $\mu(\varphi_n)$?
\end{question}

The above question seems very difficult but the next one might be more approachable. 

\begin{question}
Can the proofs we employed in the case of the $(m, n)$-transfer axioms be extended to show that 
the Lemmon-Scott's axioms, $\pd^m\Box^i p \to \Box^j\pd^n p$, are absolutely minimal among those defining the 
first-order condition 
\[
xR^my\wedge xR^jz\to \exists t(yR^it\wedge zR^nt)?
\]
\end{question}
An (admittedly weak) indication that the answer to the second question might be ``yes'' is the
fact that a slight modification of some of our  frames and models 
can be used to establish   Theorem~\ref{thrm:symmetry} below, whose proof is presented in
 Appendix~\ref{sect:symmetry}.

\begin{theorem}\label{thrm:symmetry} The formula $p\to\Box\pd p$ is absolutely minimal
among the $\lang_{\pd}$-formulas that define symmetry.
\end{theorem}


\vfill

\pagebreak

\begin{appendix}

\section{Properties of the formula-complexity game on models}\label{apSGames}

We have seen that a closed game tree $T$ induces a formula $\psi_T$.
As we see next, we can also turn formulae into game trees.

\begin{lemma}\label{lemPhiStrat}
Let $\cls A$, $\cls B$ be classes of models and $\varphi \in \languni$ be so that $\cls A \models \varphi$ and $\cls B \models \neg \varphi$. Then Hercules has a winning strategy for the $(\languni, \langle\cls{A},\cls{B}\rangle)$-{\fsgm} so that any game terminates on a closed game tree $T$ with $\psi_T = \varphi$.
\end{lemma}

\proof
We proceed by induction on the structure of $\varphi$.
\smallskip

\noindent {\sc $\varphi$ is a literal.} If $\varphi$ is a literal $\iota$, then
Hercules plays the literal-move by choosing $\iota$ and the game tree $T$ is closed with $\psi_T = \iota$, as required.
\smallskip

\noindent {\sc $\varphi$ equals $\bot$.} If $\varphi$ is $\bot$, then Hercules plays the $\bot$-move and (as $\cls B$ must be empty) the game tree $T$ is closed with $\psi_T = \bot$, as required.
\smallskip

\noindent {\sc $\varphi$ is of the form $\varphi_1 \vee \varphi_2$.} Hercules can play the $\vee$-move and add two nodes $\eta_1$, $\eta_2$ labelled by $\langle\cls{A}_1,\cls{B}\rangle$ and $\langle\cls{A}_2,\cls{B}\rangle$, respectively,
where $\cls{A}=\cls{A}_1 \cup \cls{A}_2$,  $\cls{A}_1\models \varphi_1$ 
and  $\cls{A}_2\models \varphi_2$. Applying the induction hypothesis to each sub-game, for $i\in \{1,2\}$
 Hercules has a strategy for the $(\lang_{\pd}, \langle\cls{A}_i,\cls{B} \rangle)$-{\fsgm} with resulting closed game tree $T_i$ so that $\psi_{T_i} = \varphi_i$.
 This yields a closed game tree $T$ for the original game with $\psi_T =\varphi$, as desired.\smallskip

\noindent {\sc $\varphi$ is of the form $\Diamond \theta$.}  For each 
 $\pointed a \in \cls{A}$, Hercules chooses a pointed model from 
$\Box \pointed a$  that satisfies $\theta$ and collects all these pointed models in the class $\cls{A}_1$. 
Hydra replies by choosing a subset of $\Box \pointed b$ for each $\pointed b \in \cls{B}$ and 
collects these pointed models  in  $\cls{B}_1$. 
A new node $\eta$ labelled with $\langle\cls{A}_1, \cls{B}_1 \rangle$ is added to the game tree  as a successor to the one labelled with $\langle\cls{A},\cls{B}\rangle$.
 It is obvious that  $\cls{A}_1\models \theta$ and $\cls{B}_1\models \neg \theta$. 
Applying the  induction hypothesis,  we conclude that Hercules has a strategy for the sub-game starting at $\eta$  so that the resulting game tree $S$ is closed with $\psi_S = \theta$. This yields a closed tree $T$ for the original game with $\psi_T =\Diamond\theta$.
\smallskip

\noindent {\sc $\varphi$ of the form $\exists \theta$.} 
The proof of this case follows the lines of that of $\Diamond \theta$.
 \smallskip

\noindent{\sc other cases:} Each of the remaining cases is dual to one discussed above and we omit it.
\endproof

Next we show that if the Hydra plays greedily, then any closed game tree $T$ for the $(\languni, \langle\cls{A},\cls{B}\rangle)$-{\fsgm} is such that $\cls A \models \psi_T$ and $\cls B \models \neg \psi_T$. 

\begin{lemma}\label{lemGreed}
Let $\cls A$, $\cls B$ be classes of models and let $T$ be a closed game tree for 
the $(\languni, \langle\cls{A},\cls{B}\rangle)$-{\fsgm} on which the Hydra played greedily. 
Then, $\cls A \models \psi_T$ and $\cls B \models \neg \psi_T$.
\end{lemma}

\proof
For a node $\eta$ of $T$, let $T_\eta$ be the subtree with root $\eta$, and let $\psi_\eta = \psi_{T_\eta}$. By induction on the size of $T_\eta$ starting from the leaves we show that $\lft (\eta) \models \psi_\eta$ and $\rgt (\eta) \models \neg \psi_\eta$.
The base case is immediate since Hercules can only play a literal when it is true on the left but false on the right, and inductive steps for $\bot$, $\top$, $\vee$ and $\wedge$ are straightforward. The critical case is when Hercules plays a modality on $\eta$, which is when we use that the Hydra plays greedily. 
For a $\pd$-move on $\eta$ with daughter $\eta'$, 
for each $\pointed l \in \lft(\eta)$ he chose $\pointed l' \in \Box \lft (\eta)$ and placed $\pointed l' \in \lft(\eta')$; by the induction hypothesis $\pointed l'\models \psi_{\eta'}$, so that by the semantics of $\pd$, $\pointed l \models \pd \psi_{\eta'} =\psi_\eta$. Meanwhile for $\pointed r \in \rgt (\eta)$, if $\pointed r' \in \Box \pointed r$ then since the Hydra played greedily $\pointed r' \in \rgt (\eta')$, and since $\pointed r'$ was arbitrary we see that $\pointed r \models \neg\Diamond \psi_{\eta'}$. The case for a $\Box$-move is symmetric and the cases of the $\exists$- and $\forall$-moves are analogous.
\endproof

With this we prove Theorem \ref{thrm: satisfactionGames}. 
Recall that Theorem \ref{thrm: satisfactionGames} states that the following are equivalent:
\begin{enumerate}

\item\label{itThmSatOne} Hercules has a winning strategy for the $(\languni, \langle\cls{A},\cls{B}\rangle)$-{\fsgm} with $\mu$ below $m$, and

\item\label{itThmSatTwo} there is an $\languni$-formula $\varphi$ with $\mu(\varphi) < m$ 
 and $\cls{A}\models\varphi $ whereas $\cls{B}\models\neg\varphi$.
 \end{enumerate}

\proof
Let $\cls A$, $\cls B$ be classes of models, $\mu $ any complexity measure, and $m \in \mathbb N$. 
 
 First assume that \eqref{itThmSatOne} holds, and let Hydra play the $(\languni, \langle\cls{A},\cls{B}\rangle)$-{\fsgm} greedily. By using his winning strategy, Hercules can ensure that the game terminates on some closed tree $T$ with $\mu(T) < m$. But by definition this means that $\mu(\psi_T) < m$, and by Lemma \ref{lemGreed}, $\cls A \models \psi_T$ while $\cls B \models \neg \psi_T$.

Conversely, if \eqref{itThmSatTwo} holds, by Lemma \ref{lemPhiStrat} Hercules has a strategy so that no matter how the Hydra plays, any match ends with a closed tree $T$ with $\psi_T = \varphi$, so that in particular $\mu(T) < m$.
\endproof

\section{The symmetry axiom}\label{sect:symmetry}
This appendix contains the proof of Theorem~\ref{thrm:symmetry}, that is, we show that the formula $\overline{p}\vee\Box\pd p$
is absolutely minimal  among the $\lang_{\pd}$-formulas defining symmetry.

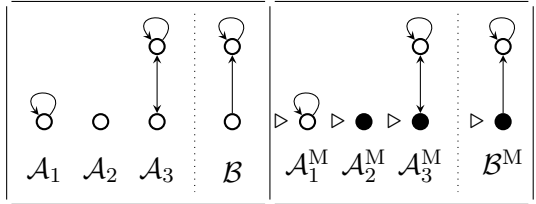
\begin{figure} [h]
\begin{center}

\begin{tikzpicture}
[
help/.style={circle,draw=white!100,fill=white!100,thick, inner sep=0pt,minimum size=1mm},
white/.style={circle,draw=black!100,fill=white!100,thick, inner sep=0pt,minimum size=2mm},
black/.style={circle,draw=black!100,fill=black!100,thick, inner sep=0pt,minimum size=2mm},
]


\node at (0.5,0) [help](A1) [label=90: $\frm{A}_1$]{};
\node at ( 0.5,1) [white](loopA1) [ ]{};

\begin{scope}[>=stealth, auto]

\draw [->] (loopA1) [out=50, in=130, loop]to  (loopA1);

\end{scope}

\node at (1.25,0) [help](A2) [label=90: $\frm{A}_2$]{};
\node at ( 1.25,1) [white](A2) [ ]{};

\node at (2,0) [help](A3) [label=90: $\frm{A}_3$]{};
\node at ( 2,1) [white](A31) [ ]{};
\node at ( 2,2) [white](A32) [ ]{};
\begin{scope}[>=stealth, auto]

\draw [<->] (A31) to  (A32);
\draw [->] (A32) [out=50, in=130, loop]to  (A32);
\end{scope}

\node at ( 2.5,2.5) [help](helpUpDotted) []{};
\node at ( 2.5,0) [help](helpDownDotted) []{};
\begin{scope}[>=stealth, auto]



\draw [ dotted] (helpUpDotted) to  (helpDownDotted);


\end{scope}


\node at (3,0) [help](B) [label=90: $\frm{B}$]{};
\node at ( 3,1) [white](rootB) [ ]{};
\node at ( 3,2) [white](loopB) [ ]{};

\begin{scope}[>=stealth, auto]

\draw [->] (rootB) to  (loopB);
\draw [->] (loopB) [out=50, in=130, loop]to  (loopB);

\end{scope}

\node at ( 3.5,-0.1) [help](mhelpDown) []{};
\node at ( 3.5,2.6) [help](mhelpUp) []{};
\node at ( 0,2.6) [help](mhelpleftUp) []{};
\node at ( 0,-0.1) [help](mhelpleftDown) []{};
\node at ( 7,2.6) [help](mhelpRightUp) []{};
\node at ( 7,-0.1) [help](mhelpRightDown) []{};

\begin{scope}[>=stealth, auto]

\draw  (mhelpUp) to  (mhelpDown);
\draw  (mhelpleftUp) to  (mhelpleftDown);

\draw  (mhelpRightUp) to  (mhelpRightDown);

\draw  (mhelpUp) to  (mhelpRightUp);
\draw  (mhelpUp) to  (mhelpleftUp);
\draw   (mhelpDown) to (mhelpleftDown);
\draw   (mhelpDown) to (mhelpRightDown);
\end{scope}


\node at (4,0) [help](mA1) [label=90:  $\mods{\frm A}_1$]{};
\node at ( 4,1) [white](mloopA1) [label=180:$\triangleright$ ]{};

\begin{scope}[>=stealth, auto]

\draw [->] (mloopA1) [out=50, in=130, loop]to  (mloopA1);

\end{scope}

\node at (4.75,0) [help](mA2) [label=90:  $\mods{\frm A}_2$]{};
\node at ( 4.75,1) [black](mA2) [ label=180:$\triangleright$]{};

\node at (5.5,0) [help](mA3) [label=90:  $\mods{\frm A}_3$]{};
\node at ( 5.5,1) [black](mA31) [label=180:$\triangleright$ ]{};
\node at ( 5.5,2) [white](mA32) [ ]{};
\begin{scope}[>=stealth, auto]

\draw [<->] (mA31) to  (mA32);
\draw [->] (mA32) [out=50, in=130, loop]to  (mA32);

\end{scope}

\node at ( 6,2.5) [help](mhelpUpDotted) []{};
\node at ( 6,0) [help](mhelpDownDotted) []{};
\begin{scope}[>=stealth, auto]



\draw [ dotted] (mhelpUpDotted) to  (mhelpDownDotted);


\end{scope}


\node at (6.6,0.1) [help](mB) [label=90:  $\mods{\frm B}$]{};
\node at ( 6.6,1) [black](mrootB) [label=180:$\triangleright$ ]{};
\node at ( 6.6,2) [white](mloopB) [ ]{};

\begin{scope}[>=stealth, auto]

\draw [->] (mrootB) to  (mloopB);
\draw [->] (mloopB) [out=50, in=130, loop]to  (mloopB);

\end{scope}

\end{tikzpicture}
\end{center}
\caption{The sets of frames  $\cls{A}=\{{\frm A}_{1}, {\frm A}_{2}, {\frm A}_{3}\}$ and $\cls{B}=\{{\frm B}\}$ and the 
respective models based on them.}
\label{fig:symmetry}
\end{figure}

Let us consider a $( \lang_{\pd},\langle \cls{A},\cls{B}\rangle)$-{\fsgf} where
$\cls{A}=\{\frm{A}_1, \frm{A}_2, \frm{A}_{3}\}$ while $\cls{B}$ contains a single element $\frm{B}$, as depicted in the left rectangle in Figure~\ref{fig:symmetry}.
Note that all frames in $\cls{A}$ are symmetric whereas this is not true about the frame $\frm{B}$.
Hence the formula $\overline{p}\vee\Box\pd p$  is valid on the frames in $\cls{A}$ and not valid on $\frm{B}$. 
\medskip

\noindent{\sc selection of the models on the right:}
Using Lemma~\ref{lemA1B}, we see that Hercules must pick the pointed model $(\mods{\frm{B}}, \triangleright)$
 shown in the right half of  Figure~\ref{fig:symmetry}. Again, to indicate that the two points of $\mods{\frm{B}}$ satisfy different sets of literals, we   
colour  one of them black and the other white.
\medskip

\noindent{\sc selection of the pointed models on the left:}
The Hydra replies as shown on the left of the dotted line in the right half in Figure~\ref{fig:symmetry}. 
Recall, that points  satisfying the same set of literals  have  the same colour.
\medskip

\noindent{\sc formula size game on models:}
Now we consider the {\fsgm} starting with  
$\Mods{\cls A} = \{ (\mods{\frm A}_1, \triangleright), (\mods{\frm A}_{2}, \triangleright), (\mods{\frm A}_{3}, \triangleright)\}$ on the left and $\Mods{\cls B} = \{(\mods{\frm B}, \triangleright)\}$
on the right.

\begin{lemma}\label{lem:RecVee}
 In any closed game tree $T$ for the $( \lang_\pd , \langle \Mods{\cls A},\Mods{\cls B} \rangle )$-{\fsgm}, Hercules played at least one $\vee$-move and at least one $\Box$-move.
\end{lemma}

\proof
The proof is almost identical to the proof of Lemma~\ref{lemRecVee}.
Indeed, it is immediate from  Lemma~\ref{lem:A1Bbox},  that if Hercules wants to win the game,
 he must not play either a $\pd$- or a $\Box$-move at
a  position $\eta$ in which $( \mods{\frm{A}}_1, \triangleright)$ is on the left and $(\mods{\frm B}, \triangleright)$ is on the right.
On the other hand, for every game in which  $(\mods{\frm A}_2, \triangleright)$ is among the pointed models chosen by the Hydra 
 and $(\mods{\frm B}, \triangleright)$ is among the models chosen by Hercules, if he wants to win the game, then  there is at least one game position  $\nu$ such that   $(\mods{\frm A}_2, \triangleright)$ is on the left and $(\mods{\frm B}, \triangleright)$ is on the right and Hercules played a $\Box$-move at $\nu$.
This implies that in any  {\fsgm} with a starting position in which the pointed models selected by the Hydra
 are on the left and $(\mods{\frm B}, \triangleright)$ is on the right, Hercules must play at least one $\vee$
to separate $(\mods{\frm{A}}_1, \triangleright)$ from  $(\mods{\frm A}_2, \triangleright)$
and one subsequent $\Box$-move.
\endproof
\begin{lemma}\label{lem:symmetrydiamond}
 In any closed game tree $T$ for the $(\lang_\pd, \langle \Mods{\cls A},\Mods{\cls B} \rangle )$-{\fsgm} where the Hydra played greedily, 
 Hercules played at least one $\pd$-move.
\end{lemma}

\proof
Let us consider a game position $\eta$ with $(\mods{\frm A}_3, \triangleright)$ on the left and $(\mods{\frm B}, \triangleright)$
on the right and let us suppose that Hercules attempts to win the  {\fsgm} with $\eta$ as a starting position
 without playing a $\pd$-move.
Clearly, a literal move is impossible at $\eta$. By playing a $\vee$- or a $\wedge$-move, he will arrive to at least
one new position that is essentially the same as $\eta$. If he plays a $\Box$-move he must select the successor of 
$(\mods{\frm B}, \triangleright)$ based on the reflexive white point in $\mods{\frm B}$. The Hydra is going to reply with the successor of  
 $(\mods{\frm A}_3, \triangleright)$ based on the reflexive white point in $\mods{\frm A}_3$. It is immediate 
that in this new game position a literal move is impossible; moreover,  no amount of $\Box$-moves are going to help Hercules win the game. Hence, Hercules must make at least one $\pd$-move.
\endproof

Thus, Theorem~\ref{thrm:symmetry} is immediate from Lemma~\ref{lem:RecVee} and Lemma~\ref{lem:symmetrydiamond}.


\end{appendix}

\end{document}